\documentclass{elsart}

\usepackage{moreverb}
\usepackage{tikz}
\usepackage{url}
\usetikzlibrary{decorations.markings}
\usetikzlibrary{matrix,shapes,arrows,positioning,chains,shapes.geometric,calc,patterns,
decorations.pathmorphing,decorations.markings}
\usepackage{graphics}
\usepackage{graphicx}
\DeclareGraphicsExtensions{.pdf,.png}
\usepackage{epsfig}
\include{epsfig}
\usepackage{amssymb}

\usepackage{booktabs}
\usepackage{amsmath}
\usepackage{graphicx}
\usepackage{psfrag}
\usepackage{amsbsy}
\usepackage{epstopdf}
\usepackage{setspace}
\usepackage{amsfonts}
\usepackage{setspace}
\usepackage{color}
\usepackage{float}
\usepackage{caption}
\usepackage{tabularx}
\usepackage{epsfig}
\usepackage{bm}
\usepackage{dcolumn} 
\usepackage{subfig}
\usepackage{pdfpages}
\graphicspath{{figures/}} % Specifies the directory where pictures are stored
\usepackage{booktabs}
\usepackage{multirow}
\makeatletter
\long\def\@maketablecaption#1#2{\@tablecaptionsize
    \global \@minipagefalse
    \hbox to \hsize{\parbox[t]{\hsize}{\centering #1 \\ #2}}}% ADDED

\makeatother
\def\vec#1{\mbox{\boldmath $#1$}}

\usepackage[english]{babel}
\usepackage{algorithm}
\def\bx{{\vec x}}
\def\bu{{\vec u}}
\def\bw{{\vec w}}

\def\bT{{\vec T}}
\def\G{\Gamma}

\def\stress{{\vec \sigma}}
\def\Otf{\Omega^\mathrm{f}(t)}

\def\Otnm1f{\Omega^\mathrm{f}(t^{\mathrm{n}-1})}
\def\Otnm12f{\Omega^\mathrm{f}(t^{\mathrm{n}-\frac{1}{2}})}

\def\Otnm1s{\Omega^\mathrm{s}_\mathrm{i}(t^{\mathrm{n}-1})}
\def\Otnm12s{\Omega^\mathrm{s}_\mathrm{i}(t^{\mathrm{n}-\frac{1}{2}})}

\def\div{\vec \nabla}

\def\real{\mathbb R}
\def\testf{\vec \phi^\mathrm{f}}

\def\dO{\mathrm{d}\vec{\Omega}}
\def\dG{\mathrm{d}\vec{\Gamma}}

\def\usnp1{{\vec u}^\mathrm{s,n+1}}

\def\vphinp1{\vec{\varphi}^\mathrm{s,n+1}}

\def\b1{\mbox{\boldmath $1$}}

\def\bB{\mbox{\boldmath $B$}}
\def\bF{\mbox{\boldmath $F$}}
\def\bG{\mbox{\boldmath $G$}}

\def\bR{\mbox{\boldmath $R$}}
\def\bT{\mbox{\boldmath $T$}}

\def\bu{\mbox{\boldmath $u$}}
\def\bw{\mbox{\boldmath $w$}}
\def\bx{\mbox{\boldmath $x$}}
\def\bnabla{{\mbox{\boldmath $\nabla$}}}

\def\bx{{\vec x}}
\def\bx{{\vec x}^\mathrm{f}}
\def\bu{{\vec u}}

\def\bT{{\vec T}}
\def\G{\Gamma}

\def\stress{{\vec \sigma}}
\def\Otf{\Omega^\mathrm{f}(t)}

\def\Otnm1f{\Omega^\mathrm{f}(t^{\mathrm{n}-1})}
\def\Otnm12f{\Omega^\mathrm{f}(t^{\mathrm{n}-\frac{1}{2}})}

\def\Of{\Omega^\mathrm{f}}

\def\Otnm1s{\Omega^\mathrm{s}(t^{\mathrm{n}-1})}
\def\Otnm12s{\Omega^\mathrm{s}(t^{\mathrm{n}-\frac{1}{2}})}

\def\d{\mathrm{d}}
\def\real{\mathbb R}

\def\testf{{\vec \phi}^\mathrm{f}}

\def\dO{\mathrm{d}\vec{\Omega}}
\def\dG{\mathrm{d}\vec{\Gamma}}

\def\bb{\mbox{\boldmath $b$}}

\def\bd{\mbox{\boldmath $d$}}

\def\bu{\mbox{\boldmath $u$}}

\def\bw{\mbox{\boldmath $w$}}
\def\bx{\mbox{\boldmath $x$}}

\def\bt{\mbox{\boldmath $t$}}

\def\bnabla{\mbox{\boldmath $\nabla$}}

\def\bF{\mbox{\boldmath $F$}}
\def\bG{\mbox{\boldmath $G$}}

\def\bR{\mbox{\boldmath $R$}}
\def\bT{\mbox{\boldmath $T$}}

\def\Gma{\mbox{\boldmath $\Gamma$}}

\def\testf{\boldsymbol{\phi^\mathrm{f}}}

\def\dO{\mathrm{d}\boldsymbol{\boldsymbol{\Omega}}}
\def\bubar{\bar{\mbox{\boldmath $u$}}}
\def\bsbar{\bar{\mbox{\boldmath $\sigma$}}}
\def\Oe{\Omega^\mathrm{e}}
\def\Of{\Omega^\mathrm{f}}

\def\Otf{\Omega^\mathrm{f}(t)}

\newcommand{\nwc}{\newcommand}

\nwc{\qref}[1]{(\ref{#1})} %\eqref seems to have a spacing problem

\renewcommand{\div}{\nabla\!\cdot\!}
\nwc{\ip}[1]{\langle #1 \rangle}

\newcommand{\xx}{\mbox{\boldmath $x$}}

\newcommand{\bvphi}{\mbox{\boldmath $\varphi$}}

\renewcommand{\div}{\nabla \cdot}

\renewcommand{\>}{\right>}
\nwc{\ta}{\tilde{a}}

\def\bu{{\vec u}}

\def\bT{{\vec T}}
\def\G{\Gamma}

\def\stress{{\vec \sigma}}
\def\Otf{\Omega^\mathrm{f}(t)}

\def\Otnm1f{\Omega^\mathrm{f}(t^{\mathrm{n}-1})}
\def\Otnm12f{\Omega^\mathrm{f}(t^{\mathrm{n}-\frac{1}{2}})}

\def\Otnm1s{\Omega^\mathrm{s}(t^{\mathrm{n}-1})}
\def\Otnm12s{\Omega^\mathrm{s}(t^{\mathrm{n}-\frac{1}{2}})}

\def\div{\vec \nabla}

\def\real{\mathbb R}
\def\testf{\vec \phi^\mathrm{f}}

\def\dO{\mathrm{d}\vec{\Omega}}
\def\dG{\mathrm{d}\vec{\Gamma}}

\usepackage{physics}

\newcommand{\pd}[2]{\frac{\partial #1}{\partial #2}}
\newcommand{\pdtwo}[2]{\frac{\partial^2 #1}{\partial #2^2}}
\newcommand{\lpd}[2]{\partial #1/\partial #2}
\newcommand{\brac}[1]{\left( #1 \right)}

\newcommand{\ssgs}{{\boldsymbol{\sigma}}^\mathrm{sgs}}

\begin{document}
%\atxy(25mm,27mm){\includegraphics{jfs_head.eps}}
%\runningheads{R. K. Jaiman}{Nonlinear Iterative Force Correction for Vortex Induced Vibrations}
\begin{frontmatter}

%========================================================== 
% Define commands to assure consistent treatment throughout document
\title{3D Common-Refinement Method 
for Non-Matching Meshes in Partitioned Variational Fluid-Structure Analysis} 
\author{Y. L. Li,}
% Authors
\author{Y. Z. Law},
\author{V.  Joshi},
\author{R. K. Jaiman\corauthref{cor}}
\ead{mperkj@nus.edu.sg}
%\author[first]{Ravi},
%\corauth[cor]{N R Pillalamarri}
\address{Department of Mechanical Engineering, National University of Singapore, Singapore 119077 }
\corauth[cor]{Corresponding author}

\begin{abstract}
We present a three-dimensional (3D) common-refinement method for non-matching meshes 
between discrete non-overlapping subdomains of incompressible fluid and nonlinear hyperelastic structure. 
The fluid flow is discretized using a stabilized Petrov-Galerkin method, and the large deformation structural 
formulation relies on a continuous Galerkin finite element method. 
An arbitrary Lagrangian-Eulerian formulation with a 
nonlinear iterative force correction (NIFC) coupling is achieved in a staggered partitioned manner 
by means of fully decoupled implicit procedures for the fluid and solid discretizations. 
%To account for the missing effects of 
%off-diagonal Jacobian terms in the partitioned staggered scheme, 
%we construct an approximate interface force correction through sub-iterations  the interaction of 
%an incompressible fluid with a low-mass density structure.
%
To begin, we first investigate the accuracy of common-refinement method (CRM) to satisfy 
traction equilibrium condition along the fluid-elastic interface with non-matching meshes.
We systematically assess the accuracy of CRM against the matching grid solution 
by varying grid mismatch between the fluid and solid meshes over a cylindrical tubular elastic 
body. We demonstrate second-order accuracy of CRM through uniform refinements of 
fluid and solid meshes along the interface.
We then extend the error analysis to transient data transfer across non-matching meshes between 
fluid and solid solvers.
We show that the common-refinement discretization across non-matching fluid-structure grids yields 
accurate transfer of the physical quantities across the fluid-solid interface.
%and the error depends on the degree of mismatch between the fluid and solid meshes along the interface.
%
We next solve a 3D benchmark problem of a cantilevered hyperelastic plate behind a circular bluff 
body and verify the accuracy of coupled solutions with respect to the available solution in the literature.
By varying the solid interface resolution, we generate various non-matching grid ratios 
and quantify the accuracy of CRM for the nonlinear structure interacting 
with a laminar flow. We illustrate that the CRM with the partitioned NIFC treatment 
is stable for low solid-to-fluid density ratio and non-matching meshes.
Finally, we demonstrate the 3D parallel implementation of common-refinement with NIFC scheme 
for a realistic engineering problem of drilling riser undergoing complex vortex-induced vibration 
with strong added mass effects. 
\end{abstract}


\end{frontmatter}
\section{Introduction}
% General data transfer
Many scientific and engineering simulations that involve interaction of multiple physical fields
often require an accurate and conservative scheme to transfer the physical data
across non-matching discrete meshes.
These problems include electromagnetics \cite{lee2005non,peng2010non},
contact dynamics \cite{hueber2009thermo,bathe,wohlmuth}, conjugate heat transfer \cite{aerothermo}, and
fluid-structure interaction (FSI) \cite{jaiman_jcp}.
% FSI and accuracy
In particular, FSI applications generally rely on different
mesh requirements for fluid and solid subdomains to capture the interaction physics accurately, which involves multiple scales and complex multi-modal coupled dynamics.
Such requirements of non-matching meshes are common in FSI applications spanning from aircraft wings, deep-water drilling riser, mooring lines,
tendons and subsea pipelines \cite{jaiman_omae2,law2017wake,blevins1990flow}
to blood blow in arteries and various biomechanical problems.
A non-conservative data transfer and locally inaccurate interpolation and projection
may lead to poor estimation of flow-elastic response and instability prediction, especially
when the frequency is close to the natural frequency of the structure.
Therefore, high-fidelity coupled fluid-structure simulations require accurate
and conservative treatment of interface boundary conditions across non-matching surface
meshes.

%% Overview of FSI methods -> Partitioned 
For body-fitted Eulerian-Lagrangian coupling, two main approaches exist for the numerical modeling of FSI problems, namely monolithic \cite{blom,hubner,turek2006,liu2014,gurugubelli2015self} and partitioned 
\cite{felippa01,cebral1997conservative,farhat05,piperno_new,yenduri2017new}. 
In the monolithic approach, the flow and structure equations are solved together in a fully coupled 
manner by assembling the coupled equations into a single block \cite{blom,turek2006,hubner}.  
While the monolithic formulations offer good numerical
stability for problems involving very strong added mass
effects, the   schemes lack the advantage of flexibility and modularity 
of using existing stable fluid or structural 
solvers \cite{turek2006,liu2014,jaiman_ficf2015}. 
On the other hand, the partitioned schemes  solve the  fluid and structure
equations in a sequential manner over two decomposed subdomains 
\cite{jaiman_cibc,piperno_new}, 
facilitating the coupling of the existing fluid and structural codes 
with suitable choices of spatial and temporal discretizations. 
The partitioned schemes can be classified 
as either strongly-coupled \cite{matthies,ahn,jaiman2016stable} or loosely-coupled
\cite{jaiman_cibc,farhat05} and the surface boundary data must be exchanged 
or transferred through the interface meshes between the fluid and structure fields. 

% Need for non-matching meshes
In a typical partitioned-based FSI simulation, 
surface meshes at the fluid-structure interface are generally non-matching 
\cite{cebral1997conservative,farhat1998load}. 
This means that their connectivity arrangements are different, 
and their geometric coordinates may not be coincident due to 
discretization requirements. 
Such non-matching meshes and associated data transfer problems also exist 
in other situations \cite{de2007review}, 
such as adaptive meshing and multi-grid considerations. 
There can be numerous conservative and non-conservative ways to interpolate and 
project data across non-matching meshes.
The data transfer must be numerically accurate and physically conservative in 
FSI simulations, especially for those that are time-dependent. 
This is because errors may accumulate over iterations and long time integration, 
and a scheme that is both accurate and conservative tends to introduce smaller 
errors and deliver an improved convergence than non-conservative or locally inaccurate approaches. 
In our work, we focus on the 3D implementation of common-refinement scheme 
based on the weighted residual or $L_2$ minimization process \cite{jaiman2006conservative}.
There are two questions one needs to answer when dealing with the spatial coupling methods across non-matching meshes  \cite{jaiman_jcp,jaiman_ijnme}: 
(a) How to interpolate and project tractions in conservative manner across fluid-solid interface with non-matching meshes? 
(b) How to integrate the traction vector defined on fluid mesh over the interface elements of 
solid mesh with their respective shape functions? 
In \cite{jaiman_jcp,jaiman_ijnme}, a detailed survey of point-to-element and common-refinement 
based scheme was provided for one-dimensional (1D) and two-dimensional (2D) problems with 
both flat and curved boundaries. It was demonstrated that the point-to-element schemes can lead to 
significant errors and sensitivity to grid-mismatch due to a violation of regularity of 
quadrature rule. These errors can be non-convergent during simultaneous grid refinement 
of fluid and solid input meshes and can impact the local accuracy along the fluid-solid boundary.

% Focus of present paper
A common-refinement overlay mesh is a surface mesh composed of elements that subdivide the 
elements of both fluid and solid input meshes simultaneously 
\cite{jiao2004common,jiao2004overlaying}, or simply the intersections of the 
elements of the input meshes. 
%A common-refinement defines and allows efficient 
%query of a unique nearby corresponding point on the fluid side for every point on the structure side and vice-versa. 
The common-refinement discretization enables accurate integration of functions that depend on the shape 
functions of the two meshes \cite{jaiman_jcp,jaiman_ijnme}. Defined as the topological intersection
of the source and target meshes, consistency of the integrations is obtained by performing the
numerical  quadrature over  the  common-refinement overlay surface.   
The common-refinement overlay mesh is constructed such that both  the  source  functions  and  the  target  functions 
are continuous in each of its elements, yielding both accuracy and conservation via accurate integration  \cite{jaiman_ijnme,slattery2016mesh}.
For these reasons, the common-refinement scheme is important for stable, accurate 
and conservative computations of fluid-structure interaction. 
This scheme is deemed desirable, but can be somewhat complex 
to implement in three dimensions as compared to other simple nearest 
neighbor or point-to-element projection methods. 
Therefore, it has been avoided by application scientists and engineers in their 
coupled partitioned multiphysics analysis. 
Indeed it is challenging to compute a common-refinement surface, since the geometrical realizations of 
the meshes are defined by distinct surfaces with arbitrary mesh intersections. 
Detailed computational geometry issues related to the construction of common-refinement 
of two three-dimensional discrete surfaces with curvature and sharp features are 
provided in \cite{jiao2004overlaying,jiao2004overlaying2}.

% Objectives
The objectives of this paper are two folds. The first is to quantify the error 
introduced by 3D common-refinement and compare against the matching reference 
counterpart.  The scope of the present work is to remove the necessity 
of matching meshes and to redesign a more general projection scheme for non-matching 
fluid and solid nodes along curved three-dimensional surface. 
Earlier investigations in  \cite{jaiman_ijnme,jaiman_jcp} were performed in one- and two-dimensional configurations and the fluid flow was considered to be compressible.
As shown in \cite{brum1,jaiman_jam} for a model elastic plate, the added mass of a compressible flow system has a dependency on the length of time interval, whereas the added mass of an 
incompressible system asymptotically approaches a constant value as the length of the time interval goes to zero.  This fundamental difference in the behavior of compressible and incompressible flows has an implication on the 
design of partitioned staggered algorithms \cite{jaiman2016stable,forster,nobile} and therefore it 
is worth investigating the stability of common-refinement scheme for an incompressible flow interacting 
with an elastic structure.
Furthermore, in the earlier works \cite{jaiman_ijnme,jaiman_jcp}, the common-refinement method was implemented for 
finite volume fluid and finite element solid solvers. In the present contribution, we employ 
the common-refinement interface between two consistent Galerkin-based variational 
formulations for fluid and solid subdomains. 
We assess the accuracy and convergence of 3D common-refinement for a circular cylinder 
tube problem in both static and transient situations.

The second objective of this work is to extend 
our NIFC implementation \cite{jaiman2016stable} with the common-refinement scheme 
for large-scale problems in FSI simulation.
The proposed computational framework integrates an ALE-based 
filtered Navier-Stokes solver, an implicit nonlinear hyperelastic structure solver, 
and the common-refinement scheme with nonlinear iterative force correction 
\cite{jaiman2016stable,jaiman2016partitioned}.
While the nonlinear structure model is discretized using 
a continuous Galerkin (CG) finite element discretization, 
a fluid solver using Petrov-Galerkin finite element spatial discretization and semi-discrete time stepping 
has been considered for the incompressible fluid flow.  
The temporal discretization of both the fluid and the structural equations is embedded in the
generalized-$\alpha$ framework by employing the classical 
Newmark approximations in time \cite{jaiman2016stable}.
Owing to domain decomposition strategy in the partitioned iterative procedure, 
we independently construct the three-dimensional meshes for the fluid and the solid subdomains.
The forces from the fluid are applied to the
structural boundary as surface tractions, and the structure displacements give a 
deformation of the fluid subdomain. The fields are advanced explicitly and the 
interface force correction is constructed at the end of each fluid sub-iteration.
%Through the fully implicit scheme, we not only achieve the coupled 
%fluid-structure stability for low solid-to-fluid mass ratio 
%but also the coupled solver is stable for a large time step size.
During the nonlinear sequence transformation, approximate interface force corrections 
are dynamically formed through sub-iterations to satisfy the force equilibrium while maintaining 
the velocity continuity condition along the fluid-solid interface.
This iterative sequence coupling relies on the generalized Aitken's iterated $\Delta^2$ process 
and the dynamic sequence parameter, which provides a fluid-structure stability at 
low structure-to-fluid mass ratio \cite{jaiman2016stable}.
We demonstrate the applicability of the new variational formulation based on the hybrid CRM-NIFC technique
for laminar and turbulent flows and compare against the reference solutions.
Finally, we demonstrate the 3D non-matching FSI computational framework for the 
vortex-induced vibration (VIV) prediction of long flexible cylinder 
in a viscous incompressible flow with strong added mass effects.

%% Organization
The organization of this manuscript is as follows. 
Section \ref{sec:numsolver} summarizes the flow and structural governing equations with interface coupling conditions. 
Section \ref{sec:spaceDiscrete} gives the spatial 
discretization of the governing equations and the interface coupling conditions. 
Discretization details of common-refinement scheme and NIFC-based coupling are 
described in Section \ref{sec:CRscheme}. 
A systematic study on the spatial accuracy of 
the common-refinement scheme is presented in Section \ref{sec:errorAnalysis}, which is followed 
by the demonstration of the accuracy of common-refinement scheme for the FSI benchmark of 
cylinder-foil problem in Section \ref{sec:validation}. 
Section \ref{sec:riser} gives a realistic engineering application of long flexible riser 
using the coupled framework based on the common-refinement and NIFC schemes. 
Finally, the work is concluded with some key findings in Section \ref{sec:conclusion}. 

%%%%%%%%%%%%%%%%%%%%%%%%% Governing equations %%%%%%%%%%%%%%%%%%%%%%%%%%%
%%%%%%%%%%%%%%%%%%%%%%%%% Numerical methodology %%%%%%%%%%%%%%%%%%%%%%%%%%%
\section{Governing fluid-structure equations}
\label{sec:numsolver}
Before the presentation of 3D common-refinement scheme,
we provide for completeness a brief review of the fluid-structure system.
The governing equations for the fluid  are applied in
an arbitrary Lagrangian Eulerian form while the dynamical structural equation is 
formulated in a Lagrangian way, and the interface conditions are enforced between 
the two physical fields.
\subsection{Incompressible Navier-Stokes with ALE formulation}
To simulate the interaction of incompressible fluid flow with a flexible structure, the body-fitted moving boundary based approach is considered in this study.
Let $\Otf \subset \real^{\mathrm{d}}$ be a fluid subdomain at time
$t$, where $\mathrm{d}$ is the space dimension. 
The motion of an incompressible viscous fluid in $\Otf$ is
governed by the Navier-Stokes equations given by 
\begin{align}
\rho^\mathrm{f}\frac{\partial \bubar^\mathrm{f}}{\partial t} \bigg\rvert_{\widehat{x}} + \rho^\mathrm{f}\left(\bubar^\mathrm{f}-\bw^\mathrm{f}\right)\cdot \boldsymbol{\nabla} \bubar^\mathrm{f} &= \boldsymbol{\nabla} \cdot \overline 
{\boldsymbol{\sigma}}^\mathrm{f} + \boldsymbol{\nabla} \cdot {\boldsymbol{\sigma}}^\mathrm{sgs} + \boldsymbol{b}^\mathrm{f}, &&\mbox{ on } \Otf, \label{eq:NS} \\ 
\boldsymbol{\nabla} \cdot\bubar^\mathrm{f} &= 0, &&\mbox{ on } \Otf, \label{eq:continuity}
\end{align}
where $\bubar^\mathrm{f}=\bubar^\mathrm{f}(\bx^\mathrm{f},t)$ and $\bw^\mathrm{f}=\bw^\mathrm{f}(\bx^\mathrm{f},t)$ represent the fluid and mesh velocities respectively, defined for each spatial fluid point $\bx^\mathrm{f} \in  \Otf$, $\rho^\mathrm{f}$ is the density of the fluid and $\boldsymbol{b}^\mathrm{f}$ is the body force applied on the fluid and $\ssgs$ represents the extra stress term due to the subgrid filtering procedure for large eddy simulation. Here,  $\bsbar^\mathrm{f}$ is the Cauchy stress tensor for a Newtonian fluid, written as
 $\bsbar^\mathrm{f} = -\bar{p}^\mathrm{f} \boldsymbol{I} + \mu^\mathrm{f}(\boldsymbol{\nabla} \bubar^\mathrm{f} + \left(\boldsymbol{\nabla} \bubar^\mathrm{f})^T\right)$,
where $\bar{p}^\mathrm{f}$ represents the filtered fluid pressure, $\mu^\mathrm{f}$ is the dynamic viscosity of the fluid. The spatial and temporal coordinates are represented  by $\bx^\mathrm{f}$ and $t$, respectively. The first term in Eq.~(\ref{eq:NS}) represents the partial derivative 
of $\bar{\boldsymbol{u}}^\mathrm{f}$ with respect to time while the ALE 
referential coordinate $\hat{x}^\mathrm{f}$ is kept fixed.
Based on the formulation in \cite{jaiman2016partitioned}, the filtered Navier-Stokes equations (\ref{eq:NS}-\ref{eq:continuity}) in the weak form can be written as 
\begin{align}
\int_{\Otf} \rho^\mathrm{f}\left(\partial_t\bubar^\mathrm{f}
+\left(\bubar^\mathrm{f}-\bw^\mathrm{f}\right)\cdot\boldsymbol{\nabla}\bubar^\mathrm{f}\right)&\cdot\testf (\xx) \dO +
\int_{\Otf} (\overline \stress^\mathrm{f} + \stress^\mathrm{sgs}) : \boldsymbol{\nabla}\testf (\xx) \dO 
\notag \\ 
\vspace{-2cm} = \int_{\Otf} &\vec{{b}}^\mathrm{f} \cdot \testf (\xx) \dO+\int_{\G^\mathrm{f}_\mathrm{h}(t)} \vec{\mathrm{h}}^{\mathrm{f}} \cdot \testf (\xx) \dG, \label{eq:WeakFormNavierStokes1}  \\
\int_{\Omega^\mathrm{f}(t)}\div  \bubar^\mathrm{f} q (\xx) \dO = 0. \quad & \label{eq:WeakFormNavierStokes}
\end{align}
Here $\partial_t$ denotes the partial time derivative operator ${\partial \left( \boldsymbol{\cdot} \right) }/{\partial t}$, $\testf$ and $q$ are the test functions for the fluid velocity and pressure, respectively. $\G^\mathrm{f}_\mathrm{h}(t)$ represents the 
non-interface Neumann boundary along which $\stress^\mathrm{f}(\bx^\mathrm{f},t) \cdot \vec{{n}}^\mathrm{f}=\vec{\mathrm{h}}^\mathrm{f}$, where $\vec{n}^\mathrm{f}$ is the normal to the fluid boundary.
The update of the deformable fluid subdomain is performed by
means of the ALE formulation \cite{hughes_ale,donea}. 
The movement of the internal finite element nodes is achieved by solving 
a continuum hyperelastic model for the fluid mesh such that the mesh quality does 
not deteriorate as the displacement of the body increases.

\subsection{Nonlinear hyperelastic structure}
We present the principle of virtual work to express the equations of motion 
and equilibrium of stresses acting on the structure. The principle of virtual work 
forms the basis for the finite element method for the dynamics of solids, which 
will be discussed later in the next section.
For a dynamically deforming structure with large strains, 
we use a nonlinear hyperelastic formulation \cite{bower2009applied} 
in the coupled fluid-structure system. 
Consider a solid with mass density $\rho^\mathrm{s}$ 
that undergoes deformation under external load by fluid flow. 
Each point on the solid is specified by its position vector. 
Let $\bx^\mathrm{s} \in \Omega_0^\mathrm{s}$ denote the initial reference position of a point in an undeformed solid, while $\bd^\mathrm{s}(\bx^\mathrm{s},t) \in \Omega^\mathrm{s}(t)$ denote the displacement of the point $\bx^\mathrm{s}$ in the deformed solid after some time $t$. 
The function $\bvphi^\mathrm{s}(\bx^\mathrm{s},t) = \bx^\mathrm{s} + \bd^\mathrm{s}(\bx^\mathrm{s},t)$ 
is thus a mapping from initial position $\bx^\mathrm{s}$ to position at time $t$, 
which completely specifies the change in shape of the solid. 
The velocity field $\bu^\mathrm{s}(\bx^\mathrm{s},t)$, which is defined as $u_i^\mathrm{s}(x_j^\mathrm{s},t) = \pd{\varphi_i^\mathrm{s}(x_j^\mathrm{s},t)}{t}$, describes the motion of the solid under the deformation. 
The external load, $\bt^\mathrm{s}$ is applied on part of the boundary of 
the solid. We use $\Gamma^\mathrm{s}_2$ to denote the solid boundary that is subjected 
to external force (Neumann boundary condition) at the reference 
configuration $\Omega_0^\mathrm{s}$, $\Gamma^\mathrm{s}_1$ to denote the 
rest of the boundary (Dirichlet boundary condition). 
Both $\Gamma^\mathrm{s}_1$ and $\Gamma^\mathrm{s}_2$ form the boundary $\Gamma^\mathrm{s}$, 
such that $\Gamma^\mathrm{s}_1 \cup\Gamma^\mathrm{s}_2 =\Gamma^\mathrm{s}$. 
By using the principle of virtual work \cite{bower2009applied}, the following weak form is obtained:
% and need to be satisfied by the velocity field on $\Omega_s(t)$
\begin{align}
 \displaystyle{\int_{\Omega^\mathrm{s}_0}} \tau^\mathrm{s}_{ij}\delta L^\mathrm{s}_{ij} d\Omega
 - \displaystyle{\int_{\Omega^\mathrm{s}_0}} \rho^\mathrm{s} b^\mathrm{s}_i \delta{v^\mathrm{s}_i} d\Omega
 + \displaystyle{\int_{\Omega^\mathrm{s}_0}} \rho^\mathrm{s} \frac{\partial {u^\mathrm{s}_i}}{\partial{t}} \delta{v^\mathrm{s}_i} d\Omega - \displaystyle{\int_{\Gamma^\mathrm{s}_2}} t_i^\mathrm{s} \delta{v^\mathrm{s}_i} \eta^\mathrm{s} d \Gamma = 0 \label{eq:hyperweak} 
\end{align}
%\quad  \varphi^\mathrm{s}_i = \varphi_i^* \ \mbox{ on }   \Gamma_1^\mathrm{s}, \nonumber  \\
%\sigma^\mathrm{s}_{ij} n_i = t^\mathrm{s}_j \ \mbox{ on }   \Gamma_2^\mathrm{s} \nonumber 
Here, $\tau^\mathrm{s}_{ij} = J^\mathrm{s} \sigma^\mathrm{s}_{ij}$ is the Kirchhoff stress,  
$J^\mathrm{s} = \det(\bF^\mathrm{s})$ is the Jacobian of deformation gradient tensor $\bF^\mathrm{s}$; 
$\sigma^\mathrm{s}_{ij}$ is the Cauchy stress; $\delta{L^\mathrm{s}_{ij}} = \lpd{\delta{v^\mathrm{s}_i}}{\varphi^\mathrm{s}_j}$ 
is the virtual velocity gradients, which satisfies $\delta{v^\mathrm{s}_i} = 0$ along boundary $\Gamma_1^\mathrm{s}$; 
$b^\mathrm{s}_i$ is the body force per unit mass, 
$\delta{v^\mathrm{s}_i} = \delta{v^\mathrm{s}_i(\bx^\mathrm{s})}$ is virtual velocity field; 
$\eta^\mathrm{s}$ is an inverse surface Jacobian which relates the boundary surface of $\Omega^\mathrm{s}(t)$ and $\Omega^\mathrm{s}_0$ as 
$\eta^\mathrm{s}_{ij} = \lpd{x_i^\mathrm{s}}{\xi_j}$, where $\xi_j$ is the isoparametric coordinate.
Note that the usual summation on $i,j,k,l$ are considered in Eq. \eqref{eq:hyperweak}. 
The Cauchy stress $\sigma^\mathrm{s}_{ij}$ is related to the left Cauchy-Green stress via the neo-Hookean constitutive law as
\begin{align}
	\sigma^\mathrm{s}_{ij} = \frac{\mu^\mathrm{s}}{(J^\mathrm{s})^{5/3}} ( B^\mathrm{s}_{ij}-\frac{1}{3}B^\mathrm{s}_{kk}\delta_{ij}) + K^\mathrm{s}(J^\mathrm{s}-1)\delta_{ij}, \quad \mbox{with} \quad B^\mathrm{s}_{ij} = F^\mathrm{s}_{ik} F^\mathrm{s}_{jk}
\label{eq:hypercauchyStress}
\end{align}
where $\bB^\mathrm{s}$ is the Cauchy-Green tensor, $\mu^\mathrm{s}$ and $K^\mathrm{s}$ are the shear modulus and the bulk modulus of the solid respectively, and the deformation gradient 
tensor $\bF^\mathrm{s}$ corresponding to a given displacement field is given as
\begin{equation}
F^\mathrm{s}_{ij}=\delta_{ij}+\pd{d_i^\mathrm{s}}{x_j}.
\label{eq:displacementfield}
\end{equation}
%=\delta_{ij}+\sum_{a=1}^n \frac{\partial{N^a}}{\partial{x_j}}\varphi_i^a. 
This completes the description of the hyperelastic structure used in our fluid-structure formulation.

\subsection{Coupling interface boundary conditions}
Here we present a short description of the coupling interface conditions for the 3D FSI problem which consists of a fluid domain $\Omega^\mathrm{f}(0)$, a solid domain $\Omega^\mathrm{s}_0$, and a common
interface boundary $\Gamma^\mathrm{fs}(t) = \partial \Omega^\mathrm{f}(t) \cap \partial \Omega^\mathrm{s}(t)$. 
For simplicity we only consider the external load through fluid flow as 
the solid Neumann boundary, i.e. $\Gamma^\mathrm{s}_2=\Gamma^\mathrm{fs}$.
Two interface boundary conditions corresponding to the
continuity of tractions and velocities must be satisfied along $\Gamma^\mathrm{fs}(t)$. 
Let $\Gamma^\mathrm{fs} \equiv \partial \Omega^\mathrm{f}(0) \cap \partial \Omega_0^\mathrm{s}$ be the fluid-solid interface at $t=0$ and $\Gamma^\mathrm{fs}(t) =\bvphi^\mathrm{s}(\Gamma^\mathrm{fs},t)$  be the interface at time $t$. The required conditions to be satisfied are as follows:
\begin{align}
	\bar{\boldsymbol{u}}^\mathrm{f}(\bvphi^\mathrm{s}(\boldsymbol{x}^\mathrm{s},t),t) &= \boldsymbol{u}^\mathrm{s}(\boldsymbol{x}^\mathrm{s},t),\\
	\int_{\bvphi^\mathrm{s}(\gamma,t)} \boldsymbol{\sigma}^\mathrm{f}(\boldsymbol{x}^\mathrm{f},t)\cdot \boldsymbol{n} \mathrm{d\Gamma}(\boldsymbol{x}^\mathrm{f}) + \int_{\gamma} \bt^\mathrm{s} \mathrm{d\Gamma} &= 0
\end{align}
where $\bvphi^\mathrm{s}$ denotes the position vector mapping the initial position $\boldsymbol{x}^\mathrm{s}$ of the flexible body to its position at time $t$, $\bt^\mathrm{s}$ is the fluid traction acting on the body, and  {$\boldsymbol{u}^\mathrm{s}$ is the structural velocity at time $t$ given by $\boldsymbol{u}^\mathrm{s} = \partial\vec\varphi^\mathrm{s}/\partial t$}. Here, $\boldsymbol{n}$ is the outer normal to the fluid-body interface, $\gamma$ is any part of the interface $\Gamma^\mathrm{fs}$ in the reference configuration, $\mathrm{d\Gamma}$ denotes the differential surface area and $\bvphi^\mathrm{s}(\gamma,t)$ is the corresponding fluid part at time $t$. The above conditions are satisfied such that the fluid velocity is exactly equal to the velocity of the body along the interface. The motion of the flexible body is governed by the fluid forces which includes the integration of pressure and shear stress effects on the body surface. 
%%%%%%%%%%%%%%%%%%%%%%%%%%%%%%%%%%%%%%%%%%%%%%%%%%%%%%%%%%%%%%%%%%%%%%%%%%%%%%%%%%%%%%%
\section{Partitioned variational fluid-structure system}\label{sec:spaceDiscrete}
For the sake of completeness, we next present the discretization using a stabilized  variational procedure with equal order interpolations for velocity and pressure. 
The coupled equations are presented in the semi-discrete  variational form
for the turbulent fluid flow interacting with a large deformation hyperelastic solid. 
For a partitioned treatment of the coupled fluid-structure interaction problems, 
the coupled system is independently discretized with the aid of
suitable and desired types of formulations for fluid and structural subdomains, the interface conditions associated with the force equilibrium 
and no-slip conditions. 
\subsection{Petrov-Galerkin finite element for fluid flow}
By means of finite element method, the fluid spatial domain $\Of$ is discretized into several non-overlapping finite elements $\Oe$,
$\mathrm{e} = 1, 2, ... , n_\mathrm{el}$, where $n_\mathrm{el}$ is the total number of elements. In this paper
we adopt a generalized-$\alpha$ method to integrate in time $t \in [t^\mathrm{n},t^\mathrm{n+1}]$, which can be unconditionally stable as well as second-order accurate for linear problems simultaneously. 
The scheme enables user-controlled high frequency damping, which is desirable and useful
for a coarser discretization in space and time. 
This scheme is implemented by means of a single parameter called the spectral radius $\rho_{\infty}$ which is able to dampen the spurious high-frequency responses and retain the second-order accuracy. With the aid of the generalized-$\alpha$ parameters $(\alpha^\mathrm{f}, \alpha_\mathrm{m}^\mathrm{f})$, the expressions employed in the variational form for the flow equation are given as \cite{jansen}:
\begin{align}
	\bar{\boldsymbol{u}}_\mathrm{h}^\mathrm{f,n+1} &= \bar{\boldsymbol{u}}_\mathrm{h}^\mathrm{f,n} + \Delta t\partial_t\bar{\boldsymbol{u}}_\mathrm{h}^\mathrm{f,n} + \gamma^\mathrm{f} \Delta t( \partial_t\bar{\boldsymbol{u}}_\mathrm{h}^\mathrm{f,n+1} - \partial_t\bar{\boldsymbol{u}}_\mathrm{h}^\mathrm{f,n}), \\
	\bar{\boldsymbol{u}}_\mathrm{h}^\mathrm{f,n+\alpha^f} &= \bar{\boldsymbol{u}}_\mathrm{h}^\mathrm{f,n} + \alpha^\mathrm{f}(\bar{\boldsymbol{u}}_\mathrm{h}^\mathrm{f,n+1} - \bar{\boldsymbol{u}}_\mathrm{h}^\mathrm{f,n}), \\
	\partial_t\bar{\boldsymbol{u}}_\mathrm{h}^\mathrm{f,n+\alpha_m^f} &= \partial_t\bar{\boldsymbol{u}}_\mathrm{h}^\mathrm{f,n} + \alpha^\mathrm{f}_\mathrm{m}( \partial_t\bar{\boldsymbol{u}}_\mathrm{h}^\mathrm{f,n+1} - \partial_t\bar{\boldsymbol{u}}_\mathrm{h}^\mathrm{f,n})
\end{align}
where 
\begin{align}
	\alpha_\mathrm{m}^\mathrm{f} = \frac{1}{2}\bigg( \frac{3-\rho^\mathrm{f}_\infty}{1+\rho^\mathrm{f}_\infty}\bigg), \qquad
	\alpha^\mathrm{f} = \frac{1}{1+\rho^\mathrm{f}_\infty}, \qquad
	\gamma^\mathrm{f} &= \frac{1}{2} + \alpha_\mathrm{m}^\mathrm{f} - \alpha^\mathrm{f}.
\end{align}
 
Let the space of the trial solutions be denoted by $\mathcal{S}^\mathrm{h}$ and the space of test functions be $\mathcal{V}^\mathrm{h}$. The variational form of the flow equations can be written as: find $[\bubar^{\mathrm{f,n}+\alpha^\mathrm{f}}_\mathrm{h}, \bar{p}^\mathrm{n+1}_\mathrm{h}]\in \mathcal{S}^\mathrm{h}$ such that $\forall [\testf,q] \in \mathcal{V}^\mathrm{h}$:
% % % %Double Check
\begin{align}
   &\int_{\Of_\mathrm{h}(t^\mathrm{n+1})}\rho^\mathrm{f}\left(\partial_t\bubar^{\mathrm{f,n}+\alpha_\mathrm{m}^\mathrm{f}}_\mathrm{h} + (\bubar^{\mathrm{f,n} + \alpha^\mathrm{f}}_\mathrm{h} - \boldsymbol{w}^{\mathrm{f},\mathrm{n} + \alpha^\mathrm{f}}_\mathrm{h})\cdot\bnabla\bubar^{\mathrm{f,n}+\alpha^\mathrm{f}}_\mathrm{h}\right) \cdot \testf \dO \notag \\
   +&\int_{\Of_\mathrm{h}(t^\mathrm{n+1})}(\bsbar^{\mathrm{f,n}+\alpha^\mathrm{f}}_\mathrm{h} + \stress^{\mathrm{sgs,n}+\alpha^\mathrm{f}}_\mathrm{h}): \bnabla\testf  \dO \notag
   -\int_{\Of_\mathrm{h}(t^\mathrm{n+1})}\bnabla \cdot \bubar^\mathrm{f,n+\alpha^\mathrm{f}}_\mathrm{h} q \dO \notag\\
   +&\sum_{\mathrm{e}=1}^{n_\mathrm{el}} \int_{\Omega^\mathrm{e}} \tau^\mathrm{f}_m\left(\rho^\mathrm{f}(\bubar^{\mathrm{f,n}+\alpha^\mathrm{f}}_\mathrm{h}-\boldsymbol{w}^{\mathrm{f},\mathrm{n}+\alpha^\mathrm{f}}_\mathrm{h})\cdot \bnabla \testf + \bnabla q \right) 
   \cdot \left( \rho^\mathrm{f} \partial_t{\bubar}^{\mathrm{f,n}+\alpha_\mathrm{m}^\mathrm{f}}_\mathrm{h}\right. \notag \\
   &\left.+\rho^\mathrm{f} (\bubar^{\mathrm{f,n}+\alpha^\mathrm{f}}_\mathrm{h}-\boldsymbol{w}^{\mathrm{f},\mathrm{n}+\alpha^\mathrm{f}}_\mathrm{h})\cdot\bnabla\bubar^{\mathrm{f,n}+\alpha^\mathrm{f}}_\mathrm{h}
   -\bnabla \cdot \bsbar^{\mathrm{f,n}+\alpha^\mathrm{f}}_\mathrm{h} - \bnabla\cdot \stress^{\mathrm{sgs,n}+\alpha^\mathrm{f}}_\mathrm{h}
   -\bb^\mathrm{f}(t^{\mathrm{n}+\alpha^\mathrm{f}})\right) \mathrm{d}\Omega^\mathrm{e} \notag \\
   +&\sum_{\mathrm{e}=1}^{n_\mathrm{el}} \int_{\Omega_\mathrm{e}}\bnabla \cdot \testf \tau^\mathrm{f}_c \rho^\mathrm{f} \bnabla \cdot \bubar^\mathrm{f,n+\alpha^\mathrm{f}}_\mathrm{h}\mathrm{d}\Omega^\mathrm{e} \notag \\
   =&  \int_{\Of_\mathrm{h}(t^\mathrm{n+1})} \boldsymbol{b}^\mathrm{f}(t^{\mathrm{n}+\alpha^\mathrm{f}})\cdot \testf \dO
   +\int_{\Gamma^\mathrm{f}_\mathrm{h}(t^\mathrm{n+1})} \boldsymbol{h}^\mathrm{f} \cdot \testf  \mathrm{d\Gma},
   \label{eq:NS-G-alpha}
\end{align}
\noindent where the lines 3, 4 and 5 represent the stabilization terms applied on each element locally. The remaining terms and the right-hand side constitute the Galerkin terms.
The stabilization parameters $\tau^\mathrm{f}_m$ and $\tau^\mathrm{f}_c$ appearing in the element level integrals are 
the least-squares metrics, which are added to the fully discretized formulation \cite{yuri}.
The least-squares metric $\tau^\mathrm{f}_m$ for the momentum equation is defined as:
\begin{align}
   \tau^\mathrm{f}_m = \left[\left(\frac{2\rho^\mathrm{f}}{\Delta t}\right)^2
   + {(\rho^\mathrm{f})}^2  (\bubar^\mathrm{f}_\mathrm{h}-\boldsymbol{w}^\mathrm{f}_\mathrm{h}) \cdot \boldsymbol{G} (\bubar^\mathrm{f}_\mathrm{h}-\boldsymbol{w}^\mathrm{f}_\mathrm{h})   
   + C_I {(\mu^\mathrm{f} + \mu^\mathrm{t})}^2 \boldsymbol{G} : \boldsymbol{G}\right]^{-\frac{1}{2}},
   \label{eq:taudef} 
\end{align}
where $C_I$ is the constant coming from the element-wise inverse estimate, $\boldsymbol{G}$ is the size of element contravariant metric tensor and $\mu^\mathrm{t}$ is the turbulence viscosity. The contravariant metric and the least-squares metric $\tau^\mathrm{f}_c$ are defined as: 
\begin{align}
   \boldsymbol{G} = \frac{\partial \boldsymbol{\xi}^T} {\partial \bx^\mathrm{f}} \frac {\partial \boldsymbol{\xi}}{\partial \bx^\mathrm{f}}, \qquad \tau^\mathrm{f}_c = \frac{1}{ {(tr} \boldsymbol{G}) \tau^\mathrm{f}_m},
\end{align} 
where $\boldsymbol{x}$ and $\boldsymbol{\xi}$ are the space coordinate and its parametric counterpart respectively, ${tr} \boldsymbol{G}$ denotes the trace of the contravariant metric tensor. The element metric tensor $\bG$ intrinsically deals with different element topology for different mesh discretizations. 
%
%The stabilization treatment serves two purposes: it provides stability to the velocity field in convection dominated regions of the fluid domain and it circumvents the Babu$\mathrm{\check{s}}$ka-Brezzi condition, which any standard mixed Galerkin methods are required to satisfy. The definition of metric $\tau^\mathrm{f}_m$ is an important factor which provides an appropriate combination of stability and accuracy. 
%%%%%%%%%%%%%%%%%%%%%%%%%%%%%%%%%%%%%%%%%%%%%%%%%%%%%%%%%%%%%%%%%%%%%%%%%%%%%%%%%

Linearization of the variational form is carried out by Newton-Raphson technique. Let $\Delta \bubar^\mathrm{f}$ and $\Delta \bar{p}^\mathrm{f}$ denote the increment in the velocity and pressure variables. The linearized matrix form of Eq. \eqref{eq:NS-G-alpha} is written as:
\begin{align}
	\vec{M}^\mathrm{f} \Delta \bubar^\mathrm{f} + \theta^\mathrm{f} \vec{K}^\mathrm{f}  \Delta \bubar^\mathrm{f} + \theta^\mathrm{f} \vec{N}^\mathrm{f}\Delta \bubar^\mathrm{f} + \theta \vec{G} \Delta \bar{p}^\mathrm{f} &= \vec{R}_m \label{eq:fluidMatrix1} \\
	- (\vec{G}^\mathrm{f}_M)^T \Delta \bubar^\mathrm{f} - (\vec{G}^\mathrm{f}_K)^T \Delta \bubar^\mathrm{f} + \vec{C}^\mathrm{f} \Delta \bar{p}^\mathrm{f} &= R_c \label{eq:fluidMatrix2}
\end{align}
where $\vec{M}^\mathrm{f}$ is the mass matrix, $\vec{K}^\mathrm{f}$ is the diffusion matrix, $\vec{N}^\mathrm{f}$ is the convection matrix, $\vec{G}^\mathrm{f}$ is the pressure gradient operator. $\vec{G}^\mathrm{f}_M$, $\vec{G}^\mathrm{f}_K$, and $\vec{C}^\mathrm{f}$ are the contribution of mass matrix, stiffness matrix and pressure matrix for the continuity equation respectively.  $\theta^\mathrm{f} = 2 \Delta t (1+\rho^\mathrm{f}_{\infty}) / (3-\rho^\mathrm{f}_{\infty})$ is a scalar, in which $\rho^\mathrm{f}_{\infty}$ is the spectral radius that acts as a parameter to control the high frequency damping \cite{jaiman2016partitioned}. $\vec{R}_m$ and $R_c$ are the right hand residual vectors in the linearized form for the momentum and continuity equations respectively.
%%%%%%%%%%%%%%%%%%%%%%%%%%%%%%%%%%%%%%%%%%%%%%%%%%%%%%%%%%%%
\subsection{Galerkin finite element for hyperelastic structure}
Using the principle of virtual work in Eq.~(\ref{eq:hyperweak}),
we present the finite element approximation for large structural deformations.
Without the loss of generality, 
we consider a hyperelastic material for the structure \cite{antman_book}.
The system of structural equations is solved employing the standard Gakerkin finite element technique 
by means of isoparametric elements for curved boundaries.
For an element with $n$ nodes, we denote the coordinates of each node by $x_i^a$, where the superscript $a$ is an integer ranging from $1$ to $n$, the subscript $i$ is an integer ranging from 1 to 3. The displacement vector at each nodal point will be denoted by  $d^{\mathrm{s},a}_i$. The displacement field and virtual velocity field at an arbitrary 
point within the solid is specified by interpolating between the nodal values as
\begin{equation}
d^\mathrm{s}_i(\bx^\mathrm{s})=\sum_{a=1}^{n} N^a(\bx^\mathrm{s}) d_i^{\mathrm{s}, a}, \quad \delta v^\mathrm{s}_i(\bx^\mathrm{s})=\sum_{a=1}^{n} N^a(\bx^\mathrm{s}) \delta v_i^{\mathrm{s}, a}.
\label{eq:hypernodalvalues}
\end{equation}
where $\bx^\mathrm{s}$ denotes the coordinates of an arbitrary solid point in the reference 
configuration and $N^a$  denotes the shape function. 
By substituting the appropriate deformation measure, the Kirchhoff stress can be calculated.  Note that the Kirchhoff stress depends on the displacements through the deformation gradient. 
Substituting Eq. (\ref{eq:hypernodalvalues}) into the virtual work equation, we obtain the 
following variational equation for nonlinear structure:
\begin{align}
 \displaystyle{\int_{\Omega^\mathrm{s}_0}} \rho^\mathrm{s} N^b N^a\frac{\partial^2{d_i^{\mathrm{s}, b}}}{\partial{t^2}}  d\Omega +  \displaystyle{\int_{\Omega^\mathrm{s}_0}} \tau^\mathrm{s}_{ij}[F^\mathrm{s}_{pq}(d_k^{\mathrm{s}, b})]  \frac{\partial{N^a}}{\partial{x_m}}  F^\mathrm{s,-1}_{mj}  d\Omega \nonumber \\
 - \displaystyle{\int_{\Omega^\mathrm{s}_0}} \rho^\mathrm{s} b^\mathrm{s}_i N^a  d\Omega 
  -  \displaystyle{\int_{\Gamma^\mathrm{fs}}} t_i^\mathrm{s} N^a \eta^\mathrm{s} d\Gamma=0. 
\label{eq:hypervirtualwork}
\end{align}
The volume and surface integrals in the above virtual work equation are taken over the reference configuration.
Using the inverse surface Jacobian $\eta^\mathrm{s}$ the deformed configuration is mapped back to the reference configuration.
The virtual work Eq. \eqref{eq:hypervirtualwork} gives a set of $n$ nonlinear equations with $n$ unknowns due to the geometric terms associated with the finite deformations.
Notably $ \tau^\mathrm{s}_{ij}[F^\mathrm{s}_{pq}(d_i^{\mathrm{s}, a})]$ is a functional relationship, 
which relates the Kirchhoff stress through the deformation gradient. 
This nonlinear virtual work equation is solved by means of Newton-Raphson iteration 
within each time step. We next briefly summarize the linearization process 
to construct a system of equations.

Let the correction of the displacement vector 
be $\Delta \bd^\mathrm{s}$. After linearizing the virtual work equation
Eq. \eqref{eq:hypervirtualwork} 
with respect to $\Delta \bd^\mathrm{s}$, we have the following system of linear equations: 
%\begin{equation}
% M^\mathrm{s}_{ab}\d\ddot{\varphi}_i^b + K_{aibk}d\varphi_k^b+R_i^a - F_i^a   = 0,
%\label{eq:hmotion}
%\end{equation}
\begin{equation}
	\vec{M}^\mathrm{s} \Delta \ddot{\bd^\mathrm{s}} + \vec{K}^\mathrm{s} \Delta \bd^\mathrm{s} = \vec{\mathcal{R}}_\mathrm{s}   
	\label{eq:hmotion}
\end{equation}
where $\vec{M}^\mathrm{s}$ is the finite element mass matrix, $\vec{K}^\mathrm{s}$ is the finite element stiffness matrix, 
and the net force vector  $\vec{\mathcal{R}}_\mathrm{s}$, which
consists of the external force and the residual terms. 
The expressions for the mass matrix $\vec{M}^\mathrm{s}$ and stiffness matrix $\vec{K}^\mathrm{s}$ 
are given by
\begin{align}
M^\mathrm{s}_{ab} &=  \displaystyle{\int_{\Omega^\mathrm{s}_0}} \rho^\mathrm{s} N^b N^a \d\Omega\\
K^\mathrm{s}_{aibk} &=  \displaystyle{\int_{\Omega^\mathrm{s}_0}} \frac{\partial{\tau^\mathrm{s}_{ij}}}{\partial{F^\mathrm{s}_{kl}}} \frac{\partial{N^b}}{\partial{x_l}} \frac{\partial{N^a}}{\partial{x_m}}  F_{mj}^{\mathrm{s},-1}  \d\Omega  - \displaystyle{\int_{\Omega^\mathrm{s}_0}} \tau^\mathrm{s}_{ij} \frac{\partial{N^a}}{\partial{x_m}} F_{mk}^{\mathrm{s},-1}  \frac{\partial{N^b}}{\partial{x_p}} F_{pj}^{\mathrm{s},-1}  \d\Omega \nonumber \\
& \quad - \displaystyle{\int_{\Gamma^\mathrm{fs}}} t_i^\mathrm{s} N^a \frac{\partial{\eta^\mathrm{s}}}{\partial{d_k^{\mathrm{s},b}}} d\Gamma,
\label{eq:stiffness} 
\end{align}
whereas the net force vectors $\vec{\mathcal{R}}_\mathrm{s}$ is as follows
\begin{align}
\{ \vec{\mathcal{R}}_\mathrm{s} \}^a_i &= \displaystyle{\int_{\Omega^\mathrm{s}_0}} \rho^\mathrm{s} b^\mathrm{s}_i N^a \d\Omega 
  - \displaystyle{\int_{\Omega^\mathrm{s}_0}} \rho^\mathrm{s} N^b N^a \pdtwo{d^{\mathrm{s},b}_i}{t} \d\Omega\  - \displaystyle{\int_{\Omega^\mathrm{s}_0}} \tau^\mathrm{s}_{ij} \frac{\partial{N^a}}{\partial{x_m}} F_{mj}^\mathrm{s,-1} d\Omega \notag \\
& \quad  + \underbrace{\displaystyle{\int_{\Gamma^\mathrm{fs}}} t_i^\mathrm{s} N^a \eta^\mathrm{s} d\Gamma}_{\text{External fluid force, } \vec{f}^\mathrm{s}}.
\label{eq:force}
\end{align}
Through the inverse surface Jacobian $\eta^\mathrm{s}$, we can relate the nominal $\vec{t}^\mathrm{s}_0$ and true 
traction $\vec{t}^\mathrm{s}$  as  $\vec{t}^\mathrm{s}_0 = \eta^\mathrm{s} \vec{t}^\mathrm{s}$  on the reference boundary $\Gamma^\mathrm{fs}$. 
After the current time step solution is obtained,
we integrate these equations with respect to time to update the solution in time.  
For our fluid-structure problems, we consider implicit Newmark time integration 
method to handle this dynamical nonlinear system. The stiffness matrix is 
rebuilt at each iteration within time-stepping.
In the spirit of partitioned treatment and domain decomposition, 
the fluid and solid subdomains are solved iteratively.
We next present the coupled fluid-structure matrix formulation and 
the iterative force correction procedure.

\subsection{Interface force correction scheme}
In this section, we present the coupled matrix form of the variational 
finite element equations defined in the previous subsection at the semi-discrete level 
for non-overlapping decomposition of two subdomains of fluid and structure.
A variational problem of fluid-structure system discretized by a finite element 
method gives a coupled linear system of equations with the unknowns of fluid and structure 
in the form $\vec{A} \vec{U} = \vec{\mathcal{R}}$, where \vec{\mathcal{R}} is a given right-hand side 
and \vec{U} is the vector of unknowns for the fluid-structure system. 
Corresponding to the domain decomposition, 
the set of degrees of freedom (DOF) is decomposed into  interior DOFs for the 
fluid-structure system and interface DOFs for the Dirichlet-to-Neumann (DtN) map.
Using the coupled fluid-structure Eqs. (\ref{eq:fluidMatrix1}), (\ref{eq:fluidMatrix2}),
(\ref{eq:hmotion}) and the Dirichlet-to-Neumann map
along the interface, the resultant block decomposition of the linear 
system can be expressed in the following abstract form: 
\begin{singlespace}
\begin{align}
\left[ \begin{array}{cccc} 
 \vec{A}_{ss} & 0 	     & 0            & \vec{A}_{fs}  \\
 \vec{A}_{ds} & \mathbb{I} & 0            & 0   	    \\
 0	      & \vec{A}_{dq} & \vec{A}_{qq} & 0  	     \\
 0            & 0 	     &  \vec{A}_{fq} &   \vec{A}_{ff} \\
 \end{array} \right]
 \left\{ \begin{array}{c} 
  \Delta \vec{d}^\mathrm{s} \\ 
  \Delta \vec{d}^\mathrm{I} \\ 
  \Delta \vec{q}^\mathrm{f}	 \\ 
  \Delta \vec{f}^\mathrm{I} 
  \end{array}  \right\}
= \left\{ \begin{array}{c}  
   \vec{\mathcal{R}}_s \\  
   \vec{\mathcal{R}}_I^D \\ 
   \vec{\mathcal{R}}_q \\  
   \vec{\mathcal{R}}_I^N 
   \end{array}  \right\} \label{eq:NIFC1}
\end{align}
\end{singlespace}
where $\vec{d}^\mathrm{s}$ is structural displacement,
$\vec{q}^\mathrm{f} = (\bubar^\mathrm{f},\bar{p}^\mathrm{f})$ denotes the fluid unknown variables, $\vec{d}^\mathrm{I}$ and $\vec{f}^\mathrm{I}$ are the displacement and force along the coupling interface. On the other side,
$\vec{\mathcal{R}}_s$ and $\vec{\mathcal{R}}_q = (\vec{R}_m,R_c)$ are the right hand side of the corresponding solid and fluid equations; 
$\vec{\mathcal{R}}_I^D$ and  $\vec{\mathcal{R}}_I^N$ are
the residual errors representing the imbalances during the enforcement of 
the Dirichlet (kinematic) condition and Neumann (dynamic) condition between 
the non-overlapping decomposed fluid and solid subdomains. 
The left-hand side matrix $\vec{A}$ represents the derivatives of the fluid, solid, interface equations with respect to their state variables. 
The subscripts $s, q$ denote the interior DOFs for the solid and fluid 
and $d, f$ represent the interface DOFs for the displacement and force.
While the block matrix $\vec{A}_{ss}$ corresponds to the mass and stiffness 
matrix of the structure in Eq. \eqref{eq:hmotion}, $\vec{A}_{qq}$ 
corresponds to the coupled fluid velocity and pressure linear system 
in Eq. \eqref{eq:fluidMatrix1} and \eqref{eq:fluidMatrix2}. 
$\vec{A}_{ds}$ is an extraction matrix which maps the solid displacement 
to the interface, $\vec{A}_{df}$ is a matrix which relates the displacement 
of the interface to the fluid through ALE. 
$\vec{A}_{fq}$ is the computation of the force and its mapping to the interface. 
$\vec{A}_{fs}$ is a matrix that gets the solid load vector from 
the fluid-solid interface force. 
During the partitioned Dirichlet-to-Neumann coupling,
there is no explicit availability of the Jacobian matrix $\vec{A}_{fs}$ entered 
in the coupled fluid-structure system Eq. ~\eqref{eq:NIFC1}.

By eliminating the off-diagonal term $\vec{A}_{fs}$ in Eq. \eqref{eq:NIFC1} 
via static condensation, we can obtain the following reduced linear system:
 \begin{singlespace}
\begin{align}
\left[ \begin{array}{cccc} 
 \vec{A}_{ss} & 0 	     & 0            & 0  \\
 \vec{A}_{ds} & \mathbb{I} & 0            & 0   	    \\
 0	      & \vec{A}_{dq} & \vec{A}_{qq} & 0  	     \\
 0            & 0 	     &  0 & \widetilde{\vec{A}}_{ff}  \\
 \end{array} \right]
 \left\{ \begin{array}{c} 
  \Delta \vec{d}^\mathrm{s} \\ 
  \Delta \vec{d}^\mathrm{I} \\ 
  \Delta \vec{q}^\mathrm{f}	 \\ 
  \Delta \vec{f}^\mathrm{I} 
  \end{array}  \right\}
= \left\{ \begin{array}{c}  
   \vec{\mathcal{R}}_s \\  
   \vec{\mathcal{R}}_I^D \\ 
   \vec{\mathcal{R}}_q \\  
   \widetilde{\vec{\mathcal{R}}}_I^N
   \end{array}  \right\} \label{eq:NIFC2}
\end{align}
\end{singlespace}
where the derivation and the terms $\widetilde{\vec{A}}_{ff}$ and $\widetilde{\vec{\mathcal{R}}}_I^N$ 
are described in detail in \cite{jaiman2016partitioned}. 
The idea is to compute the iterative correction for the interface fluid force 
$\vec{f}^I=\int_{\Gamma^\mathrm{fs}} \boldsymbol{\sigma}^\mathrm{f}\cdot \boldsymbol{n} \mathrm{d\Gamma}$ 
over the deformed ALE configuration as the structure moves.
In the nonlinear iterative force correction, 
we construct a force correction vector in the following manner to correct 
the previous force $\vec{f}^I_{k}$ at $k^\mathrm{th}$ sub-iteration:
\begin{equation} 
\Delta \vec{f}^I = 
\widetilde{\vec{A}}^{-1}_{ff} \widetilde{\vec{\mathcal{R}}}_{I}^{N}
\end{equation}
%\begin{equation} 
%\vec{f}^I = \vec{f}^I_{0} +
%\widetilde{\vec{A}}^{-1}_{ff} \widetilde{\vec{\mathcal{R}}}_{I}^{N}
%\end{equation}
Here, the force correction vector 
$\widetilde{\vec{A}}^{-1}_{ff} \widetilde{\vec{\mathcal{R}}}_{I}^{N}
$ at the $k^\mathrm{th}$ sub-iteration can be constructed by 
successive approximation, which essentially estimates the
coupled effects along the fluid-solid interface. 
The off-diagonal terms are not explicitly formed and the scheme 
instead proceeds in a predictor-corrector format by constructing 
an iterative interface force correction at each sub-iteration. 
This iterative force correction relies on the input-output relationship 
between the displacement and the force transfer at each sub-iteration. 
When the brute-force sub-iterations lead to severe numerical 
instabilities during strong added-mass effects, the NIFC-based 
correction provides a stability to the partitioned 
FSI coupling \cite{jaiman2016partitioned}.
The present force correction scheme can be interpreted as a generalization of 
Aitken's $\Delta^2$ extrapolation scheme \cite{breziniski2007,buoso} 
to provide convergent behavior to the interface force sequence 
generated through the nonlinear iterations between the fluid and the structure.
The geometric extrapolations with the aid of dynamic weighting 
parameter allow to transform a divergent fixed-point iteration to a 
stable and convergent iteration \cite{jaiman2016partitioned,jaiman2016stable}.
We next present the common-refinement scheme to transfer the fluid traction 
over the solid boundary across non-matching meshes.

%%%%%%%%%%%%%%%%%%%%%%%%%%%%%%%%%%%%%%%%%%%%%%%%%%%%%%%%%%%%%%%%%%%%%%%%%%
\section{3D common-refinement scheme}\label{sec:CRscheme}
In this section, we will address the central topic of this paper focusing 
on the spatial coupling between the fluid and the structure for non-matching meshes via common-refinement.
The fluid and the structural equations are coupled by the continuity of velocity and
traction along the fluid-solid interface.
We consider that the fluid-structure boundary is discretized independently by the same polynomial 
order for both fluid and solid subdomains. 
During the ALE updates, the deformed fluid subdomain using isoparametric elements 
follows the deformed structure and the discrete fluid and solid meshes remain coincident 
to the physical fluid-structure boundary.
\subsection{Variational interface condition}
To satisfy the traction equilibrium condition, the momentum flux from the fluid 
flow must be transferred to the structural surface through the surface traction.
To formulate the load transfer operator for the common-refinement method,
the weighted residual based on $L_2$ minimization is considered.
Let $N_i^\mathrm{f}$ and $N_j^\mathrm{s}$ denote the standard finite element shape functions associated
with node $i$ of the fluid and node $j$ of the solid interface meshes respectively, while $\tilde{\bold{t}}^\mathrm{f}_i \in L^2(\Omega^\mathrm{f})$ and $\tilde{\bold{t}}^\mathrm{s}_i \in L^2(\Omega^\mathrm{s})$ denote the approximate nominal tractions at the corresponding nodes of the discrete fluid interface $\Gamma^\mathrm{f}_\mathrm{h}$
and solid interface $\Gamma^\mathrm{s}_\mathrm{h}$ respectively. 
The continuum traction fields $\bold{t}^\mathrm{f}$ and $\bold{t}^\mathrm{s}$  over $\Gamma^\mathrm{f}$ and $\Gamma^\mathrm{s}$ are 
interpolated as follows:
\begin{equation}
\bold{t}^\mathrm{f}(\bx^\mathrm{f})\approx\sum_{i=1}^{m_f} N^\mathrm{f}_i\tilde{\bold{t}}^\mathrm{f}_i, \quad	 \bold{t}^\mathrm{s}(\bx^\mathrm{s})\approx\sum_{j=1}^{m_s} N^\mathrm{s}_j\tilde{\bold{t}}^\mathrm{s}_j.
\label{eq:cftractions}
\end{equation}
where $m_f$ and $m_s$ are the number of fluid and solid nodes on the fluid and solid interface meshes respectively. Once we have $\bold{t}^\mathrm{f}$, $N_i^\mathrm{f}$, and $N_j^\mathrm{s}$, we can obtain the transferred distributed loads by solving for  $\tilde{\bold{t}}^\mathrm{s}_j$. 
We can measure the residual $\bold{t}^\mathrm{s} - \bold{t}^\mathrm{f}$, by minimizing the $L_2$ norm employing the Galerkin weighted residual method. Multiplying
both sides with a set of weighting functions $(W_i =  N_i^\mathrm{s})$,  and integrating over the interface boundary $\Gamma^\mathrm{fs}$, we obtain:
\begin{equation}
\int_{\Gamma^\mathrm{fs}}N_i^\mathrm{s}\bold{t}^\mathrm{s}d\Gamma=\int_{\Gamma^\mathrm{fs}}N_i^\mathrm{s}\bold{t}^\mathrm{f}d\Gamma,
\label{eq:cf3}
\end{equation}
By using finite element approximations, the traction equilibrium condition 
is: 
\begin{equation}
\int_{\Gamma^\mathrm{fs}}N_i^\mathrm{s}N_j^\mathrm{s}\tilde{\bold{t}}^\mathrm{s}_jd\Gamma=\int_{\Gamma^\mathrm{fs}}N_i^\mathrm{s}N_j^\mathrm{f}\tilde{\bold{t}}^\mathrm{f}_jd\Gamma,
\label{eq:cf4}
\end{equation}
which gives solid-side tractions $\tilde{\bold{t}}^\mathrm{s}_j$  as
\begin{equation}
\tilde{\bold{t}}^\mathrm{s}_j=[M_{ij}^\mathrm{s}]^{-1}\{f_i^\mathrm{s}\},
\label{eq:cf5}
\end{equation}
where $[M^\mathrm{s}]$ represents the mass matrix for the solid interface elements 
and is defined by using:
\begin{equation}
[M_{ij}^\mathrm{s}]=\int_{\Gamma^\mathrm{fs}}N_i^\mathrm{s}N_j^\mathrm{s}d\Gamma,
\label{eq:cf6}
\end{equation}
and $\{f_j^\mathrm{s}\}$ is the nodal force vector given as:
\begin{equation}
\{f_i^\mathrm{s}\}=\sum_{j=1}^{m_f} \tilde{\bold{t}}^\mathrm{f}_j\int_{\Gamma^\mathrm{fs}}N_j^\mathrm{f}N_i^\mathrm{s}d\Gamma,
\label{eq:cf7}
\end{equation}
This completes the general formulation of the weighted residual method for extracting 
the load vector on the solid side interface. While the construction of mass matrix 
requires only the solid side shape functions, the load vector integral 
consists of shape functions from both the fluid and the solid sides. 
For matching grids, this would not cause any problem. However for non-matching
grids, the inconsistency of shape functions will lead to integrations across discontinuities.
To resolve this issue, there is a need for the common-refinement surface which allows 
to perform integrations consistently. Further details of common-refinement construction 
for three dimensional surface meshes can be found in \cite{jiao2004overlaying}.
\subsection{Algorithmic details}
\begin{figure}[!h]
	\centering
    \includegraphics[width=0.8\textwidth]{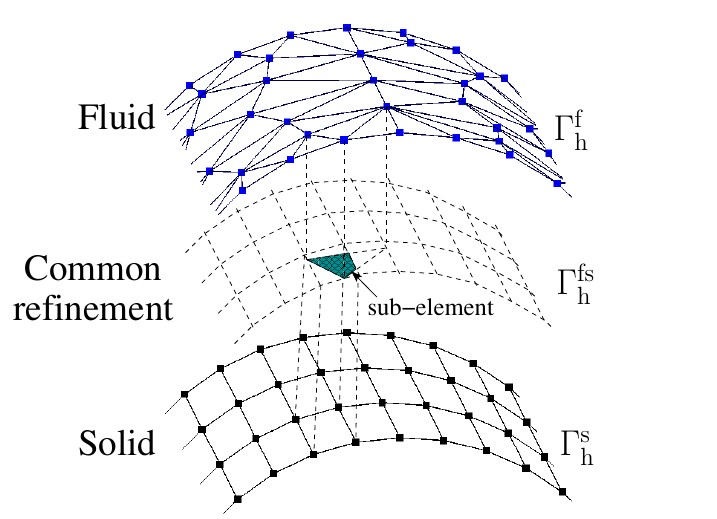}
		\caption{Schematic of common-refinement based projection scheme 
                for load transfer in 3D; where shaded area denotes one
surface sub-element. Physical fluid-solid interface has two realizations 
of fluid and solid sides.}
		\label{fig:schm_cf}
\end{figure}

The common-refinement scheme is an important and special data structure, 
for transferring data between meshes that have different mesh ratios. 
As shown in Fig.~\ref{fig:schm_cf}, 
the common-refinement surface of two meshes consists of polygons that subdivide 
the input boundary meshes of fluid and solid subdomains simultaneously. 
Every sub-element of a common-refinement mesh has two geometrical realizations, 
in general, which are different but must be close to each other, to 
obtain a physically consistent data transfer. In the finite element form, 
the spatial configuration of the fluid and solid interface meshes can be written as: 
\begin{equation}
\bx^\mathrm{f}\approx\sum_{i=1}^{m_f} N^\mathrm{f}_i(\bx)\tilde{\bx}^\mathrm{f}_i \,\, on \,\,\Gamma^\mathrm{f}_\mathrm{h}, \quad	 \bx^\mathrm{s}\approx\sum_{j=1}^{m_s} N^\mathrm{s}_j(\bx)\tilde{\bx}^\mathrm{s}_j \,\, on \,\,\Gamma^\mathrm{s}_\mathrm{h}.
\label{eq:cf10}
\end{equation}
Within this paper, the topology of the common-refinement sub-elements are defined by the 
intersection of the surface elements of input meshes. 
These 3D sub-elements are illustrated in Fig. \ref{fig:schm_cf}. 
We notice that the intersection of two arbitrary triangles or two hybrid surface elements can be quite complex. 

During FSI simulation, within the common-refinement scheme, the load vector $f_j^\mathrm{s}$ over the common-refinement 
mesh nodes is computed as follows:
 \begin{equation}
f^\mathrm{s}_j=\sum_{i=1}^{e_c}\int_{\sigma^c_i}N^\mathrm{s}_j\tilde{\bold{t}}^\mathrm{f}d\Gamma,
\label{eq:cf11}
\end{equation}
where $e_c$ represents the total number of sub-elements of the common-refinement overlay surface, and $\sigma^c_i$
represents its $i$th sub-element. We use the Gaussian integration to determine the integration point locations and their weight functions. 
The basic steps of CRM are summarized as follows:
%%%%%%%%%%%%%%%%%%%%%%%%%%%%%%%%%%%%%%%%%%%%%%%%%%%%%%%%%%%%%%%%%%%%%%%%%%%%%%%%%%%%%%%%%%%%%%
\floatstyle{ruled}
\newfloat{algorithm}{H}{loa}
\floatname{algorithm}{Algorithm}
\begin{algorithm}
\caption{3D common-refinement method}
\begin{tabbing}
1. \= Collect input meshes and construct common-refinement sub-elements \\
2.  Loop over defined sub-elements over fluid interface $\Gamma_\mathrm{h}^\mathrm{f}$  \\
\> (a) Loop over quadrature points of the sub-elements   \\

\>  (b) Calculate sub-element area over fluid interface \\
\> (c)  Associate quadrature points onto corresponding solid element \\
\> (d) Integrate traction vectors via common-refinement sub-elements \\
\> (e) Compute solid load vector integral using Eq. (\ref{eq:cf11})
\end{tabbing}
\end{algorithm}
   
\subsection{Common-refinement with NIFC scheme}
\begin{figure}[!htbp]
\centering
\begin{tikzpicture}[decoration={markings,mark=at position 0.5 with {\arrow[scale=2]{>}}},scale=1.0]
	\draw[-,black] (0,0) node[anchor=north,black]{$\Omega^\mathrm{f}$} to (0,11);
	\draw[-,black] (6,0) node[anchor=north,black]{$\Omega^\mathrm{s}$} to (6,11);
	\draw[dotted] (3,0) node[anchor=north,black]{$\Gamma^\mathrm{fs}$} to (3,11);

	\fill[black] (5.85,0.85) rectangle (6.15,1.15) ;
	\fill[black] (5.85,3.85) rectangle (6.15,4.15) ;
	\fill[black] (5.85,6.85) rectangle (6.15,7.15) ;
	\fill[black] (5.85,9.85) rectangle (6.15,10.15) ;

	\draw[black,fill=black] (0,1) circle (0.9ex);
	\draw[black,fill=black] (0,4) circle (0.9ex);
	\draw[black,fill=black] (0,7) circle (0.9ex);
	\draw[black,fill=black] (0,10) circle (0.9ex);
	
	\draw[black,postaction={decorate}] (0,1) to (6,1);
	\draw[black,postaction={decorate}] (4,0) to (6,1);
	\draw[black,postaction={decorate}] (0,10) to (2,11);

	\draw[black,postaction={decorate}] (6,1) to (6,4);
	\draw[black,postaction={decorate}] (6,4) to (0,1);
	\draw[black,postaction={decorate}] (0,1) to (0,4);
	\draw[black,postaction={decorate}] (0,4) to (6,4);

	\draw[black,postaction={decorate}] (6,4) to (6,7);
	\draw[black,postaction={decorate}] (6,7) to (0,4);
	\draw[black,postaction={decorate}] (0,4) to (0,7);
	\draw[black,postaction={decorate}] (0,7) to (6,7);

	\draw[black,postaction={decorate}] (6,7) to (6,10);
	\draw[black,postaction={decorate}] (6,10) to (0,7);
	\draw[black,postaction={decorate}] (0,7) to (0,10);
	\draw[black,postaction={decorate}] (0,10) to (6,10);

	\draw[black,thick] (-0.25,7.6) to (0.25,7.8);
	\draw[black,thick] (-0.25,7.75) to (0.25,7.95);
	\draw[white] (0,7.72) to (0,7.83) ;

	\draw[black,thick] (5.75,7.6) to (6.25,7.8);
	\draw[black,thick] (5.75,7.75) to (6.25,7.95);
	\draw[white] (6,7.72) to (6,7.83) ;

	\draw[black] (6.2,1) node[anchor=west,black]{$\bd^\mathrm{s}(\bx^\mathrm{s},t^\mathrm{n})$};
	\draw[black] (6.2,10) node[anchor=west,black]{$\bd^\mathrm{s}(\bx^\mathrm{s},t^\mathrm{n+1})$};

	\draw[black] (-0.2,1.6) node[anchor=east,black]{$\bubar^\mathrm{f}(\bx^\mathrm{f},t^\mathrm{n})$,};
	\draw[black] (-0.2,1.0) node[anchor=east,black]{$\bar{p}^\mathrm{f}(\bx^\mathrm{f},t^\mathrm{n})$,};
 	\draw[black] (-0.2,0.4) node[anchor=east,black]{$\bw^\mathrm{f}(\boldsymbol{x}^\mathrm{f},t^\mathrm{n})$};

	\draw[black] (-0.2,10.6) node[anchor=east,black]{$\bubar^\mathrm{f}(\bx^\mathrm{f},t^\mathrm{n+1})$,};
 	\draw[black] (-0.2,10) node[anchor=east,black]{$\bar{p}^\mathrm{f}(\bx^\mathrm{f},t^\mathrm{n+1})$,};
 	\draw[black] (-0.2,9.4) node[anchor=east,black]{$\bw^\mathrm{f}(\boldsymbol{x}^\mathrm{f},t^\mathrm{n+1})$};

	\draw (6,4) -- (0,1) node [midway, above, sloped] {$\bubar^\mathrm{f} = \bu^\mathrm{s} = \bw^\mathrm{f}$};
	\draw (2.3,4.3) node[anchor=west]{$\boldsymbol{f}^\mathrm{I}_{k+1} = \boldsymbol{f}^\mathrm{I}_{k} + \delta\boldsymbol{f}^\mathrm{I}$};
	
	\draw (6,2.5) node[anchor=west,black]{(1)};
	\draw (3.4,2.5) node[anchor=west,black]{(2)};
	\draw (0,2.5) node[anchor=east,black]{(3)};
	\draw (3.9,3.7) node[anchor=east,black]{(4)};

%	\draw[black,fill=white] (8,6) circle (2ex);
%	\draw[black] (8,7) node{(1)};
%	\draw (8.5,7) node[anchor=west,black]{Solve structural displacement};
%	\draw[black] (8,6) node{(2)};
%	\draw (8.5,6) node[anchor=west,black]{Transfer predicted solid displacement};
%	\draw[black] (8,4.5) node{};
%	\draw (8.5,4.5) node[anchor=west,black]{while satisfying ALE compatibility and velocity continuity at the interface};
%	\draw[black] (8,5) node{(3)};
%	\draw (8.5,5) node[anchor=west,black]{Solve ALE fluid and turbulence equations};
%	\draw[black] (8,4) node{(4)};
%	\draw (8.5,4) node[anchor=west,black]{Correct forces using NIFC filter};

	\draw (1.5,2.5) node[anchor=east]{$k=1$};
	\draw (0.35,2.25) rectangle (1.4,2.7) ;
	\draw (1.5,5.5) node[anchor=east]{$k=2$};
	\draw (0.35,5.25) rectangle (1.4,5.7) ;
	\draw (0.25,8.5) node[anchor=west]{$k=\mathrm{nIter}$};
	\draw (0.35,8.25) rectangle (2.1,8.7) ;

	\draw (6.7,3.5) node[rotate=90, anchor=west,black]{Hyperelastic structure};
	\draw (-1.0,7.5) node[rotate=90, anchor=east,black]{ALE fluid \& turbulence};
	%\draw (-0.2,5.15) node[anchor=east]{turbulence};
\end{tikzpicture}
\caption{A schematic of predictor-corrector procedure for ALE fluid-turbulence and structure coupling via nonlinear iterative force correction \cite{jaiman2016partitioned,jaiman2016stable}. (1) Solve structural displacement, (2) transfer predicted solid displacement, (3) solve ALE fluid equations, (4) correct forces using NIFC filter. Here, $\mathrm{nIter}$ denotes the maximum number of nonlinear iterations to achieve a desired convergence tolerance 
within a time step  at $ t \in [t^\mathrm{n},t^\mathrm{n+1}]$.}
% and the key algorithmic steps are: (1) Predict structural displacements, (2) Transfer computed solid displacement while satisfying ALE compatibility and velocity continuity at the interface, (3) Solve Navier-Stokes in ALE form and turbulence equations at the mid-point configurations of deformed fluid mesh, (4) Correct the forces using NIFC filter and transfer them to the structural solver. }
\label{f:nifcSchematic}
\end{figure}
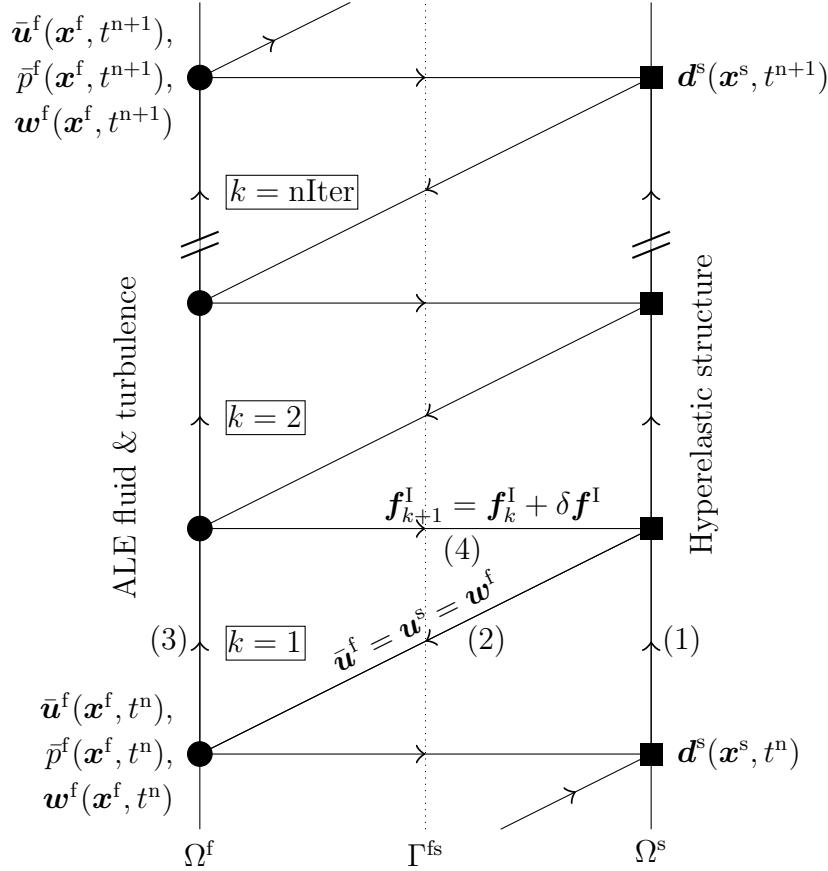

In this section, we summarize the partitioned iterative coupling of the ALE fluid-turbulence solver with the hyperelastic structure solver, as illustrated in Fig. \ref{f:nifcSchematic}. The solution to the hyperelastic structure equations provides a predictor displacement. The ALE fluid equations with turbulence are then solved to evaluate and correct the forces at the fluid-solid interface. Let the structural displacement be denoted by $\bd^\mathrm{s}(\boldsymbol{x}^\mathrm{s},t^\mathrm{n})$ due to the turbulent fluid forces at time $t^\mathrm{n}$. The first step of the iterative procedure at iteration $k$ involves the prediction of the displacement of the hyperelastic structure due to the fluid forces. The computed structural displacements are then transferred to the fluid side by satisfying the ALE compatibility and the velocity continuity conditions at the interface $\Gamma^\mathrm{fs}$ in the second step. This is elaborated as follows: suppose $\bd^\mathrm{m,n+1}$ is the mesh displacement at time $t^\mathrm{n+1}$. The mesh displacements are equated to the structural displacements at the fluid-solid interface to avoid any gaps and overlaps between the non-matching fluid and solid mesh configurations
 \begin{equation}
\bd^\mathrm{m,n+1} = \bd^\mathrm{s} \ \ \ \mathrm{on} \ \ \Gamma^\mathrm{fs}.
\label{eq:nifc1}
\end{equation}
The fluid velocity is then equated with the mesh velocity to satisfy the velocity continuity on $\Gamma^\mathrm{fs}$ as
 \begin{equation}
\bubar^\mathrm{f,n+\alpha^f} = \bw^\mathrm{f,n+1}=\frac{\bd^\mathrm{m,n+1}-\bd^\mathrm{m,n}}{\Delta t} \ \ \ \ \mathrm{on} \ \ \Gamma^\mathrm{fs}.
\label{eq:nifc2}
\end{equation}

In the third step of the iteration $k$, the ALE Navier-Stokes equations with subgrid LES filtering  are solved at the mid-point moving mesh configuration to evaluate the fluid forces. The computed forces are finally corrected using the NIFC filter and transferred to the hyperelastic structural solver in the fourth step of the nonlinear iteration. When the solver has achieved the convergence criteria, the fluid-structure solver is advanced in time after updating the variable values at $t^\mathrm{n+1}$. 
%\begin{figure}
%	\centering
%	\includegraphics[width=0.8\textwidth]{nifcSchematic.pdf}
%	\caption{A schematic of predictor-corrector procedure for ALE fluid-turbulence and structure coupling via nonlinear iterative force correction. (1) Solve structural displacement, (2) transfer predicted solid displacement, (3) solve ALE fluid and turbulence equations, (4) correct forces using NIFC filter.}
%	\label{f:nifcSchematic}
%\end{figure}

In this numerical study, we employ Newton-Raphson technique to minimize the linearization error at each time step and the flow and ALE mesh fields are updated in time by the generalized-$\alpha$ method \cite{jansen}. The resulting incremental velocity, pressure and mesh displacement coming from the finite element discretization are evaluated by solving the linear system of equations via the Generalized Minimal RESidual (GMRES) algorithm proposed in \cite{saad1986gmres}, which relies on the Krylov subspace iteration and the modified Gram-Schmidt orthogonalization. Note that to solve the linear matrix system, we do not explicitly form the left hand-side matrix, rather we perform the needed matrix-vector product of each block matrix in pieces for the GMRES algorithm. The solver relies on a hybrid parallelism for the solution 
of partitioned NIFC-based FSI solver for parallel computing. The parallelization employs a standard master-slave 
strategy for distributed memory clusters via
message passing interface (MPI) based on a domain decomposition strategy. The master process 
extracts the mesh and generates the partition of the mesh into subgrids via an 
automatic graph partitioner. 
Each master process performs the computation for the root 
subgrid and the remaining subgrids behave as the 
slaves \cite{woodsend2009hybrid,karypis1998software}. 
While the local element matrices and the local right-hand vectors are evaluated by 
the slave processes, the resulting system is solved in parallel across different compute 
nodes \cite{smith2001development}. The hybrid or mixed approach provides the benefit
of thread-level parallelism of multicore architecture and allows MPI task accessing the 
full memory of a compute node. After solving the Navier-Stokes equations, we assemble 
the tractions acting on the hyperelastic structure by means of MPI, and then transfer them onto the 
surface of the hyperelastic structure through common-refinement overlay surface.
%%%%%%%%%%%%%%%%%%%%%%%%%%%%%%%%%%%%%%%%%%%%%%%%%%%%%%%%%%%%%%%%%%%%%%%%%%%%%%%%%%%%%%%%%%%%%
\section{Error analysis and convergence study}\label{sec:errorAnalysis}
In a typical FSI simulation, the surface of fluid and solid is always intact and coincident without any gaps or overlaps. The fluid load is projected onto the solid surface, while the displacement of the solid is projected onto the fluid surface. Such data transfer is repeated over each time step. There are two types of errors that we are interested during the transferring of data. First, the error in a single transfer of data from one surface to another surface with non-matching mesh. Second, the error in a repeated transfer of data from one surface to another. The analysis for these two types of errors is conducted and described in the following sections.
\subsection{Static data transfer}
\begin{figure}[h!]
	\centering
%	\subfloat[][3D Mesh dividing with $h_s/h_f = 1.5$]{\includegraphics[width=0.55\textwidth]{error3dview15.png}} \; \; 
	\subfloat[][Presribed load transferred from fluid to solid across a cylinder surface]{\includegraphics[width=0.60\textwidth]{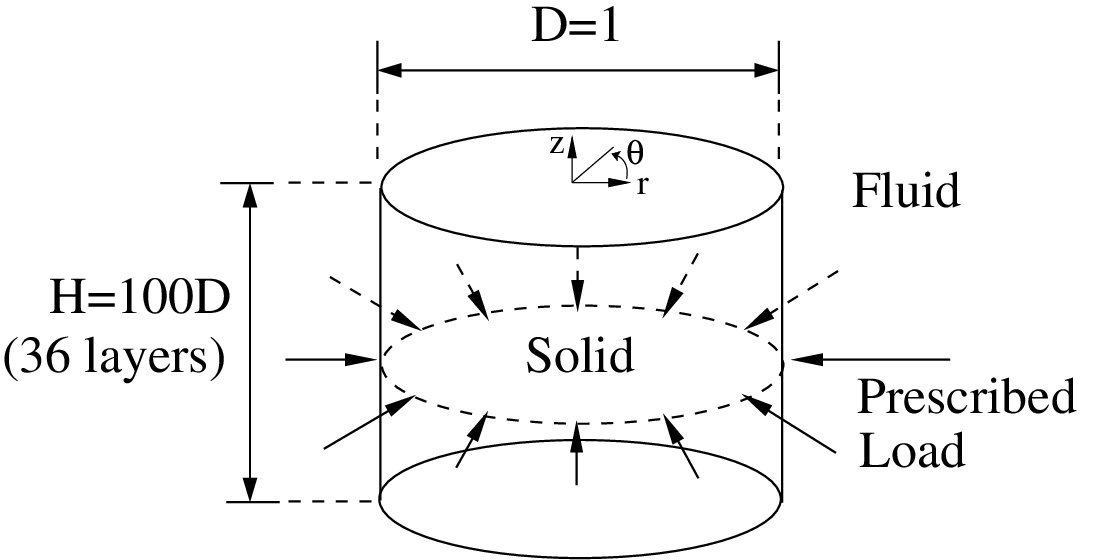} \label{figCylinderSchematic1}} \;
	\subfloat[][Typical non-matching mesh of solid (dotted line) and fluid (solid line)]{\includegraphics[width=0.35\textwidth]{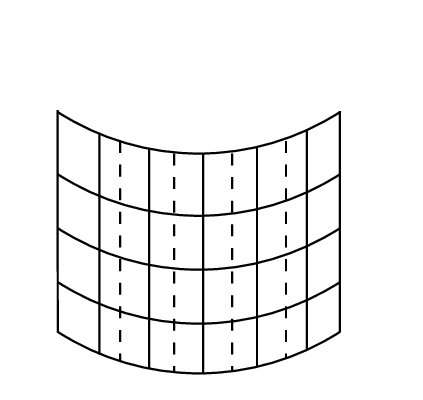}\label{figSurfaceSchematic1}}  \; \; 
	%\subfloat[][X-Y view of $h_s/h_f = 1.5$ ]{\includegraphics[width=0.45\textwidth]{error2dview15.eps}} \; \;
	%\subfloat[][X-Y view of $h_s/h_f = 0.6667$ ]{\includegraphics[width=0.45\textwidth]{error2dview06.eps}} \; \;
	\caption{ 3D non-matching FSI meshes for static error analysis of common-refinement scheme.}
	\label{fine to coarse mesh}
\end{figure}

The error introduced in a single transfer of data from one surface to another surface with non-matching mesh is investigated here. To quantify this error, we set up two intact surfaces with different mesh size. For the consistency of the notation for FSI, we define one of the surface as fluid, while the other surface as solid. Both surfaces resemble the geometry of a cylinder surface with diameter $D=1$ and height $H = 100D$, as shown in Fig. \ref{figCylinderSchematic1}. Both surfaces are discretized uniformly into $w_z$ elements in $Z$-direction. Without the loss of generality, we choose $w_z = 36$ in our analysis. Then, it is discretized into $N_s$ and $N_f$ elements along the circumference of the cylinder. The resultant number of elements on the fluid and the solid surface is thus $w_zN_f$ and $w_zN_s$ respectively. The distributions of element sizes are uniform within each surface.  Fig. \ref{figSurfaceSchematic1} shows a typical patch of the non-matching fluid and solid surface meshes. 
We define the size of the element on each surface through the area of each element as follows:
\begin{align}
	A_s = \frac{\pi D H}{w_z N_s}, \quad A_f = \frac{\pi D H}{w_z N_f}.
\end{align}
The degree of non-matching is quantified by using the mesh ratio between these two surfaces, $A_s/A_f$. Several meshes with different mesh sizes are generated by setting $N_s, N_f \in \{ 36, 54, 108, 162, 216 \}$. This results in the mesh ratio of $A_s/A_f \in [1/6,6]$. For each set of these meshes, a prescribed load is applied on the node of fluid surface mesh at position $\bx^\mathrm{f}$ and is transferred to the solid surface via common-refinement scheme. Consider the prescribed loading function in the following form:
\begin{align}
	\bt^\mathrm{s} = \bt^\mathrm{s}(\theta,z) = - \left( \frac{1}{2} \rho^\mathrm{f} U_{\infty}^2 (1-4\sin^2 \theta) + \rho^\mathrm{f} g z  \right) \begin{pmatrix}
%	\vspace{-0.5cm}
	0.5 \cos \theta \\
%	\vspace{-0.5cm}
	0.5 \sin \theta \\
	0
\end{pmatrix}	 \label{eqPrescribedLoad}
\end{align}
where $(\theta,z) \in \Gamma^\mathrm{fs}$ is the cylindrical position vector on the surface of the cylinder and $U_{\infty} = 1$. Note that the origin of the cylindrical coordinates lies at the center of the top surface of the cylinder as shown in Fig. \ref{figCylinderSchematic1}. The prescribed load in Eq. \eqref{eqPrescribedLoad} is generated through the estimation of static pressure along $Z$-direction from the potential flow around a cylinder. Let $(\theta^\mathrm{s}_j,z^\mathrm{s}_j)$ be the position vector of node $j$ on the solid's cylinder surface, $\bT^\mathrm{s}_j$ be the load vector transferred to that node,  and $|| \cdot ||_2$ be the $L_2$ norm, the relative error $\epsilon_1$ is defined as 
\begin{align}
	\epsilon_1 = \frac{\sum_{j} \norm{\bT^\mathrm{s}_j - \bt^\mathrm{s}(\theta^\mathrm{s}_j,z^\mathrm{s}_j)}_2}{\sum_{j} \norm{\bt^\mathrm{s}(\theta^\mathrm{s}_j,z^\mathrm{s}_j)}_2}. \label{eq:epsilon_1}
\end{align}
Table \ref{erroranalysis1} shows the relative error $\epsilon_1$ computed for the mesh ratio ranging from $0.1667$ to $6.0$. The error analysis shows that the common-refinement scheme performs well within the interpolant error for all the mesh ratios. More importantly, it is worth noting that the error is consistent for both $h_s/h_f > 1$ and $h_s/h_f < 1$. This is expected as the overlay surface constructed in the common-refinement scheme involves both the fluid and solid meshes.

\begin{table}[h!]
	\centering
		\caption{Dependence of load vector error $\epsilon_1$  on different mesh ratio $h_s/h_f$}
			 \label{erroranalysis1}
	\begin{tabular}{cccc}
	\toprule
	 $N_s$ & $N_f$ & $A_s/A_f$ & Error, $\epsilon_1$ \\
	 \midrule
	 36 & 216 & $6.0$ & $2.94896 \times 10^{-3}$   \\
	 36 & 162 & $4.5$ & $2.88308 \times 10^{-3}$   \\
	 36 & 108 & $3.0$ & $2.69495 \times 10^{-3}$   \\
	 36 & 54 & $1.5$ & $ 1.6836  \times 10^{-6}$   \\
	 36 & 36 & $1.0$ & $2.51 \times 10^{-16}$   \\
	 54 & 36 & $0.6667$ & $1.76919 \times 10^{-6}$ \\
	 108 & 36 & $0.3333$ & $3.30083 \times 10^{-3}$ \\
	 162 & 36 & $0.2222$ & $3.29819  \times 10^{-3}$ \\
	 216 & 36 & $0.1667$ & $3.32599 \times 10^{-3}$  \\
	 \bottomrule
	\end{tabular}
\end{table}

\begin{figure}[!h]
\centering
     \includegraphics[width=0.7\textwidth]{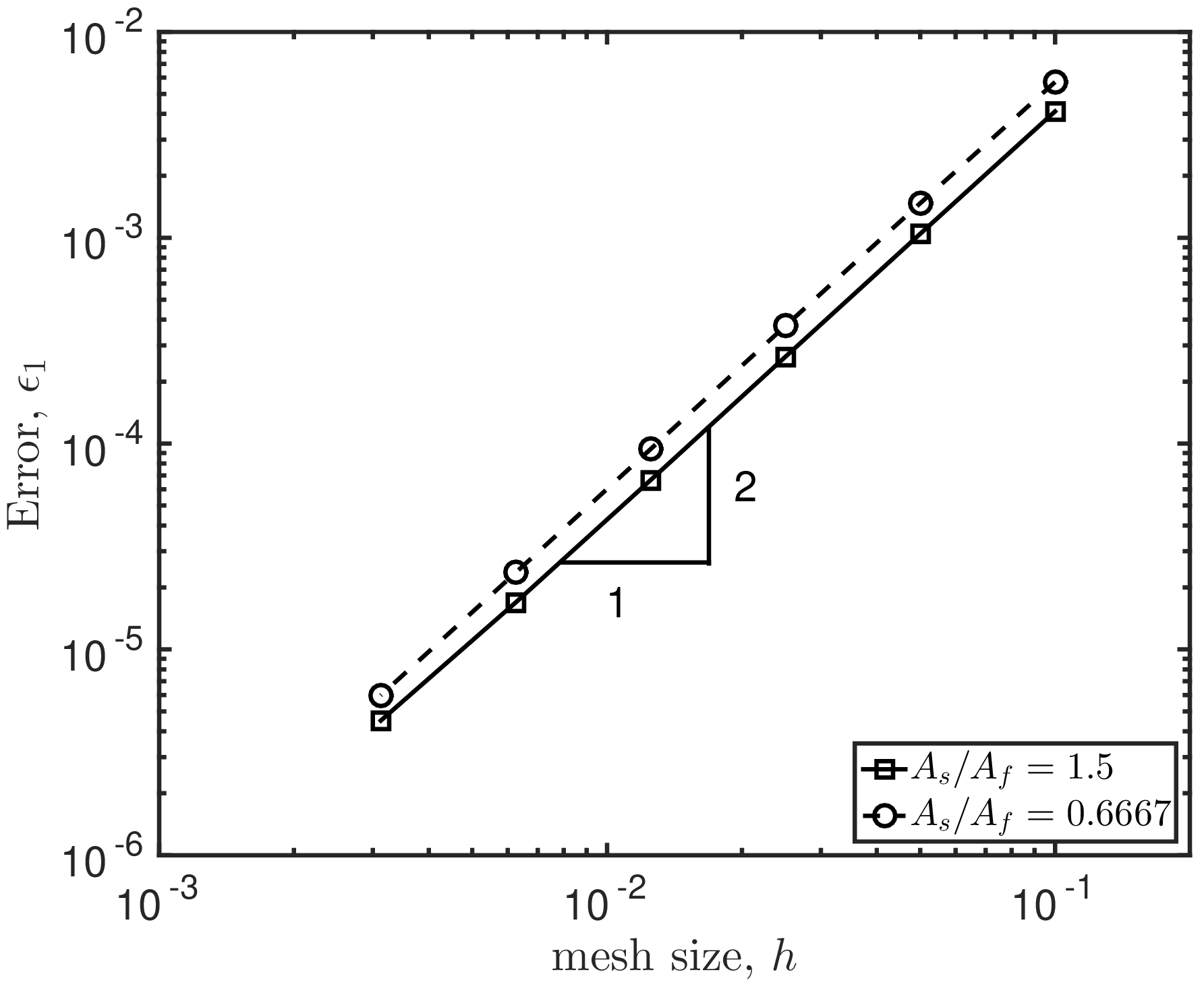}
		\caption{Spatial mesh convergence of common-refinement scheme for non-matching meshes.}
		\label{spatialerror}
\end{figure}

To further quantify the error of common-refinement scheme via spatial convergence study, we choose $h_s/h_f =1.5$ and $h_s/h_f=0.6667$ as the reference mesh ratios.
 For each mesh ratio, we increase both $N_s$ and $N_f$ simultaneously by maintaining the mesh ratio while lowering the error introduced due to spatial discretization. The relative error  $\epsilon_1$ in standard discrete least-square norm for non-matching meshes is shown in Fig.\ref{spatialerror}. It is found that the gradient of the line plotted is 2, which implies that the common-refinement scheme is optimally accurate up to the geometric interpolation. 

\subsection{Transient analysis}
\begin{figure}[h!]
	\centering
	\subfloat[][]{\includegraphics[width=0.8\textwidth]{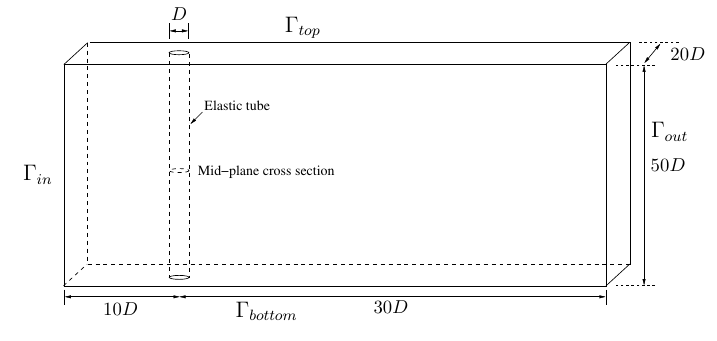} \label{figCylinderSchematic}} \\
	\subfloat[][]{\includegraphics[width=0.60\textwidth]{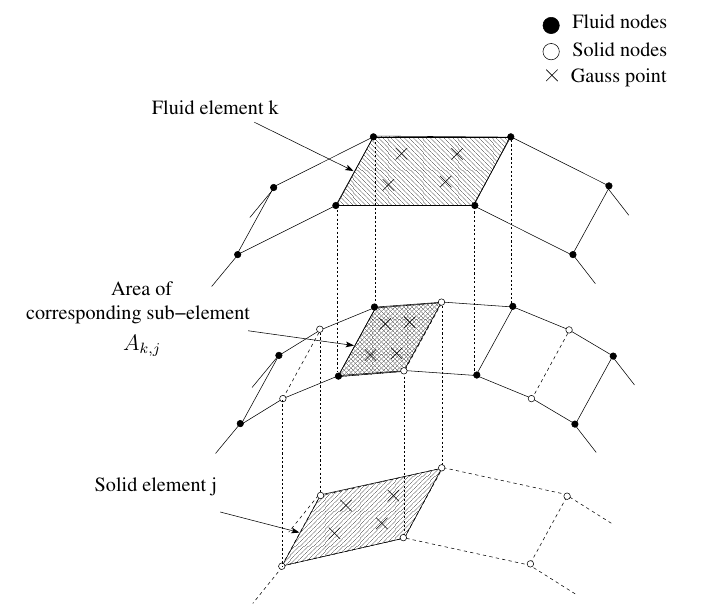} \label{figDegreeOfMismatch}} \; \;
	\begin{minipage}[c][2cm][b]{0.35\textwidth}
		\centering
		\subfloat[][]{\includegraphics[width=\textwidth]{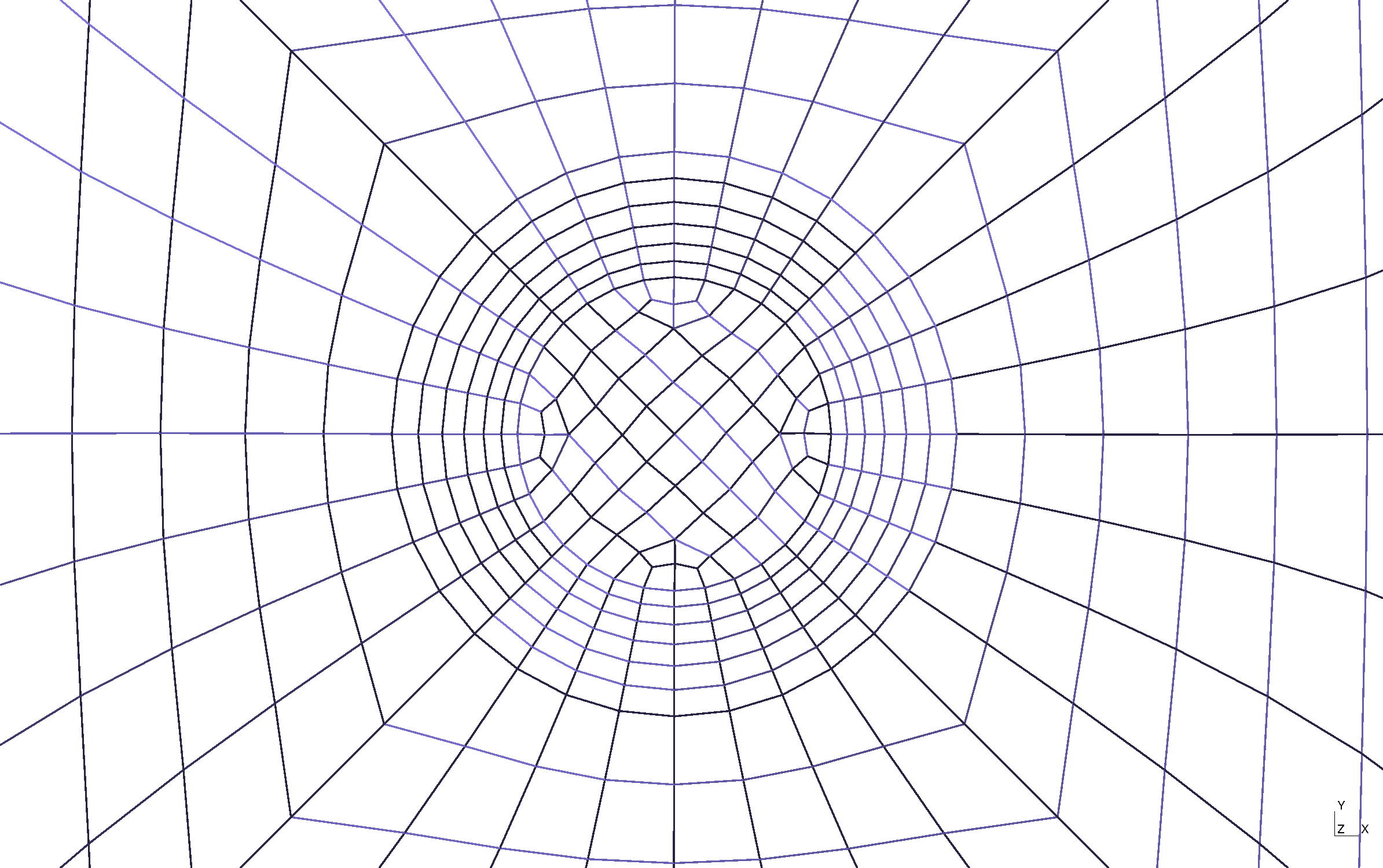}} \\
		\subfloat[][]{\includegraphics[width=\textwidth]{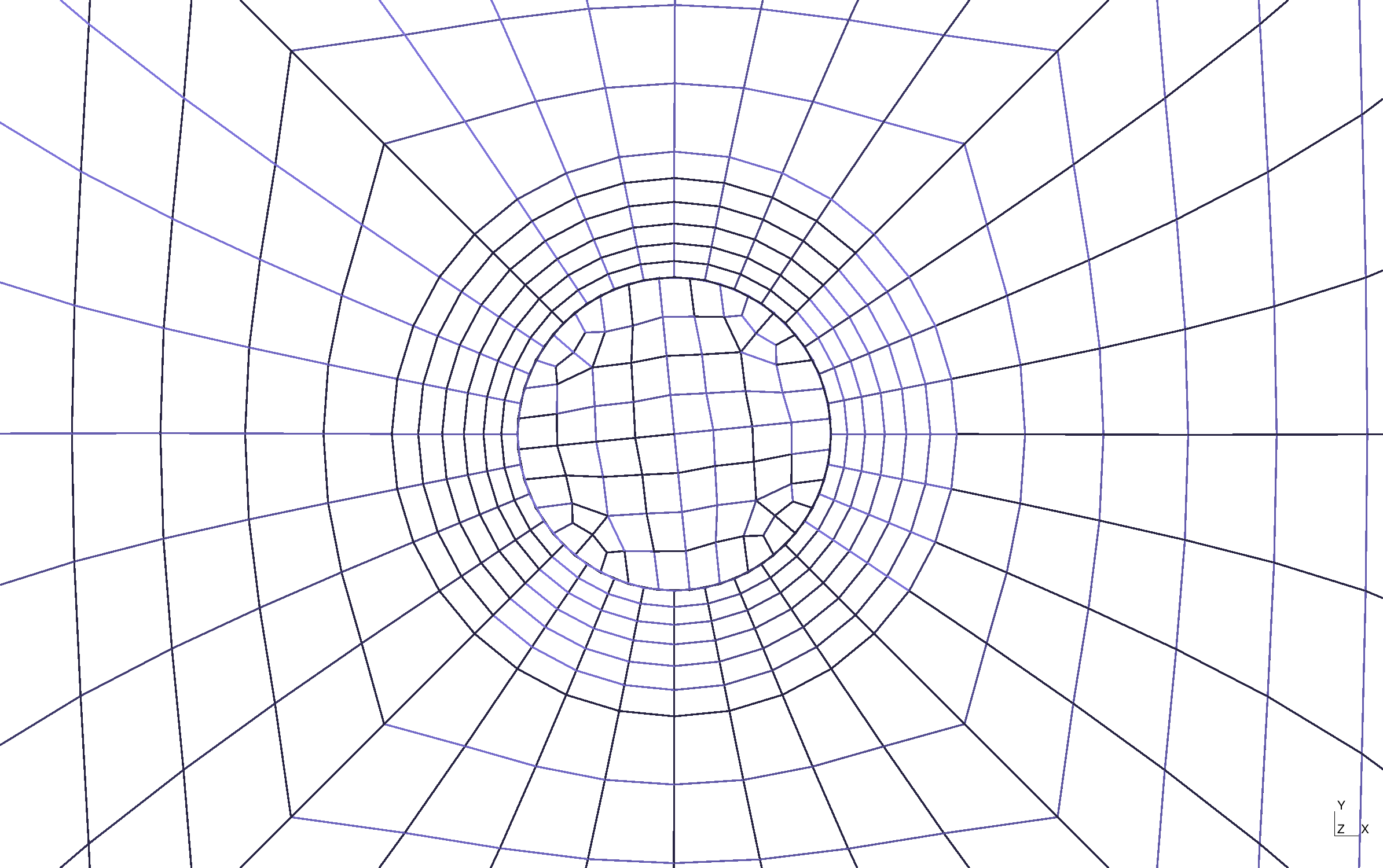}} \; \;
	\end{minipage}
	\caption{ Problem setup for the transient analysis of common-refinement discretization: (a) schematic diagram of problem setup for deformable tube problem in a uniform flow; (b) sketch of mismatching mesh and their projected area on common-refinement surface; (c) and (d) show the meshes on mid-plane cross section at matching ($\delta_{F \rightarrow S} = 0$) and non-matching ($\delta_{F \rightarrow S} = 0.5$) conditions.}
	\label{c2fmesh}
\end{figure}

The second type of error, which is introduced during the repeated transfer of data across non-matching mesh is analyzed herein. This analysis is conducted by considering a long deformable cylinder in a flowing channel for meshes with different degrees of mismatch. The schematic diagram of the problem setup is shown in Fig. \ref{figCylinderSchematic}. The cylinder is modeled as an elastic tube with spanwise length of $50D$. Both of its ends are fixed at $\Gamma_{top}$ and $\Gamma_{bottom}$. A free-stream flow velocity of $u^\mathrm{f}=U$ is specified at $\Gamma_{in}$ of the computational domain where $u^\mathrm{f}$ is the $X$-component of the fluid velocity $\bubar^\mathrm{f} = (u^\mathrm{f},v^\mathrm{f},w^\mathrm{f})$, while the traction-free boundary condition  is considered at $\Gamma_{out}$. Slip boundary condition is applied on $\Gamma_{top}$ and $\Gamma_{bottom}$, and no slip boundary condition is imposed on the surface of the cylinder. 
The cylinder is placed $10D$ away from the $\Gamma_{in}$, and $30D$ away from $\Gamma_{out}$. 

%%%%%%%%%%%%%%%%%%%%%%%%%%%%%%%%%%%%%%%%%%%%%%%%%%%%%%%%

The data transfer occurs at the surface of the flexible cylinder, where the traction of the fluid is passed to the solid. Both the fluid and the solid surfaces are discretized into $w_{\theta}$ elements along the circumference and $w_{z}$ elements along the spanwise direction. As a representative case, we choose $w_{\theta} = 32$ and $w_z=25$ in this analysis. This leads to a mesh ratio of $A_s/A_f = 1$, where the area of each fluid and solid element is the same. A mismatch between both surfaces is then generated by fixing the fluid mesh and rotating the solid mesh along cylinder's spanwise axis. Figure \ref{figDegreeOfMismatch} shows a schematic diagram for such mismatching of meshes for the fluid and the solid subdomains. For each fluid surface element $k$ and solid element $j$, there is a corresponding sub-element on the common-refinement surface. The sub-element is the intersection of the projected elements of both subdomains. Let $A_{k,j}$ be the area of the sub-element corresponding to the fluid surface element $k$ and the solid surface element $j$, the degree of mismatch with respect to the fluid element $k$, which is termed as $\delta^{F \rightarrow S}_k$ is defined as follows
\begin{align}
	\delta^{F \rightarrow S}_k = \brac{1 - \frac{\max_{j} A_{k,j}}{\sum_j A_{k,j}}}.  \label{eqDegreeOfMismatch}
\end{align}
The degree of mismatch with respect to the fluid surface $\delta^{F \rightarrow S}$ is then defined as the mean degree of mismatch with respect to each fluid element
\begin{align}
	\delta^{F \rightarrow S} = \frac{1}{N} \sum_{k=1}^N \delta^{F \rightarrow S}_k.
\end{align}
With this definition, it is found that $\delta_{F \rightarrow S} \in [0,0.5]$ in our case, where $\delta_{F \rightarrow S} = 0$ corresponds to a matching mesh, $\delta_{F \rightarrow S} = 0.5$ corresponds to a staggered configuration between the fluid and the solid elements. Five sets of meshes with different degrees of mismatch, including $\delta_{F \rightarrow S} = 0$, are used in the numerical experiments to assess the accuracy of the common-refinement scheme. The matching mesh is selected as a reference case to evaluate the error associated with the degree of mismatch. 

%To test the performance of common-refinement scheme on transient problem, we set up two cases to run a simple cylinder problem with C-R method respectively. The schematic diagram of the problem tested is shown in Fig. \ref{figCylinderSchematic}. Two mesh settings with mesh ratio = 1.0 are adopted: matching mesh and non-matching mesh in which the solid side rotated a little bit. The X-Y plane sections for different mesh setting are shown as Figure.5. 
%We adopted C-R to run matching mesh and non-matching mesh respectively. Then we compared the $C_d$, $C_l$, X-displacement and Y-displacement, to observe their behaviors in different mesh setting. The results are shown in Figure.7. 
The characteristic responses of the deformable tube are compared between the different mesh configurations. These include the in-line displacement, the cross-flow displacement, the drag coefficient ($C_d$) and the lift coefficient ($C_l$).
% as shown in Fig. \ref{figTransient}. 
%\begin{figure}[!h]
%	\centering
%	\subfloat[][X-Displacement]{\includegraphics[width=0.45\textwidth]{xdis.eps}} \; \; 
%	\subfloat[][Y-Displacement]{\includegraphics[width=0.45\textwidth]{ydis.eps}} \\
%	\subfloat[][Drag coefficient, $C_d$]{\includegraphics[width=0.45\textwidth]{cd.eps}} \; \; 
%	\subfloat[][Lift coefficient, $C_l$]{\includegraphics[width=0.45\textwidth]{cl.eps}}
%	\caption{The characteristic response of the riser with matching and non-matching mesh. It is observed that their responses are almost identical.}
%	\label{figTransient}
%\end{figure}
%It can be seen that the results of these meshes are almost identical.
To quantify the effect of degree of mismatch on these physical  quantities, we compute the error as follows:
%We compute the relative error by means of maximal norm:
\begin{equation}
\epsilon_2 = \frac{||\bR - \bR_{\delta_{F \rightarrow S}=0}||_{\infty}}{||  \bR_{\delta_{F \rightarrow S}=0} ||_{\infty}}
\label{eq:epsilon_2}
\end{equation}  
where $\bR$ is a vector containing the temporal characteristic response from the non-matching mesh, $\bR_{\delta_{F \rightarrow S}=0}$ is the corresponding temporal response of the matching mesh and $|| \cdot ||_{\infty}$ is the infinity norm. The error associated with each of the characteristic responses is summarized in Table \ref{erroranalysis2}. It can be observed that there is a very small difference between the responses during the simulation. 
Therefore, it shows that the 3D common-refinement scheme developed in this paper is both reliable and accurate in transferring the data between the fluid and the solid meshes. To further assess the stability and accuracy of our 3D fluid-structure solver with non-matching meshes, we next present a standard benchmark problem of elastic foil attached to a circular cylinder.
\begin{table}[h!]
	\centering
		\caption{The relative error of characteristic response between the matching and non-matching meshes}
		 \label{erroranalysis2}
	\begin{tabular}{ccccc}
	\toprule
	 $\delta_{F \rightarrow S}$ &  $x/D (\times 10^{-3})$ & $y/D (\times 10^{-3})$ & $C_l (\times 10^{-3})$ & $C_d (\times 10^{-3})$ \\
	 \midrule
	 0.1 & 1.2512 & 2.1213 & 0.0791 & 1.2382 \\
	 0.2 & 1.6620 & 3.0710 & 0.1102 & 1.5931 \\
	 0.3 & 3.6998 & 3.9819 & 0.1301 & 1.6318 \\
	 0.4 & 5.6904 & 4.8199 & 0.1691 & 1.8075 \\
	 0.5 & 7.5399 & 5.6292 & 0.2005 & 2.1933 \\
	 \bottomrule
	\end{tabular}
\end{table}

%%%%%%%%%%%%%%%%%%%%%%%%%%%%%%%%%%%%%%%%%%%%%%%%%%%%%%%%%%%%%%%%%%%%%%%%%%%%%%%
\section{Three-dimensional FSI with non-matching meshes}\label{sec:validation}
\begin{figure}[!h]
	\centering
	\subfloat[][Schematic of unsteady cylinder-foil problem ]{\includegraphics[width=0.90\textwidth]{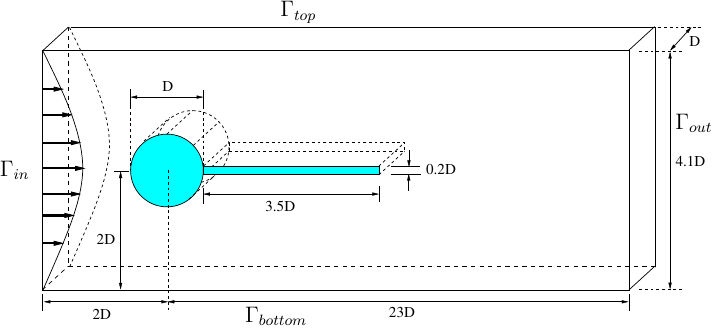} \label{figFoilSchematic}} \; \;
	\subfloat[][Representative matching mesh M1]{\includegraphics[width=0.65\textwidth]{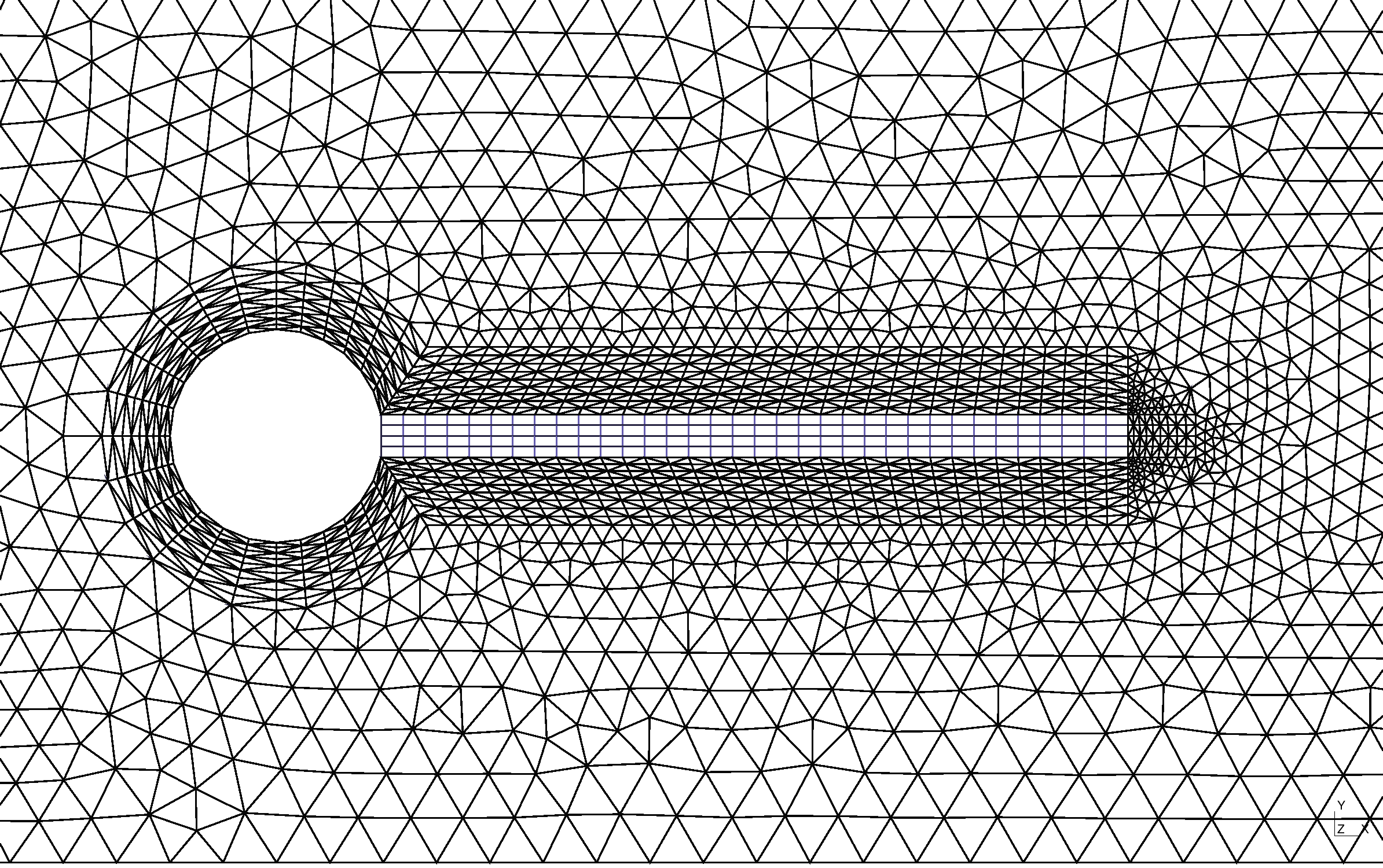} \label{figFoilMesh}} \; \; 
	\caption{ Unsteady cylinder-foil problem for the verification and 
        convergence study. Details of meshing parameters are listed in Table 4. }
	\label{Schematicbenchmark}
\end{figure}

Before we proced to the demonstration of the common-refinement scheme 
for non-matching meshes, we first verify our partitioned fluid-structure 
computation against the available data in literature. 
For this purpose, we consider the unsteady cylinder-bar problem (FSI-III) 
presented in \cite{turek2006proposal} for low structural-to-fluid density ratio and 
$Re=200$ based on the diameter of the cylinder. The simulation consists of a thin 
flexible structure with a finite thickness clamped behind a fixed 
rigid non-rotating cylinder. The cylinder-foil system is installed in 
a rectangular fluid domain. 
A schematic of the cylinder-foil system and computational domain is 
shown in Fig. \ref{figFoilSchematic}.
Table \ref{numericalpara} summarizes the fluid-structure parameters used for this 
benchmark problem.
\begin{table}[!h]
	\centering	
		\caption{FSI parameters for unsteady cylinder-foil problem 
                at $Re=200$ and $\rho^\mathrm{s}/\rho^\mathrm{f}=1.0$  }
				 \label{numericalpara}
	\begin{tabular}{cc}
		\toprule
		Parameters & Benchmark  \\
		\midrule
		Cylinder diameter , $D$       & $0.1m$  \\		
		Mean inlet velocity, $\bar{U}$		& 1.0 $m/s$		\\
		Fluid density, $\rho^\mathrm{f}$  &1000  $kg/m^3$ 	\\
%		Fluid viscosity, $\mu_f$  &$1.0\times10^{-3} $ $kgm^{-1} s^{-1}$   	%& $8.158\times10^{-3}$ $ML^{-1}T^{-1}$ 	\\
		Foil thickness, $w$ & $0.02 m$ \\
		Foil length, L & $0.35 m$ \\
		Structure density, $\rho^\mathrm{s}$  & 1000  $kg/m^3$ 	\\
		Young Modulus, $E$ 	& $5.6  \times 10^{5} Pa$ 	\\
		Reynolds number, $Re$ & 200 	 \\
		Density ratio, $\rho^\mathrm{r} = \frac{\rho^\mathrm{s}}{\rho^\mathrm{f}}$ & 1.0 \\
		Poisson's ratio, $\nu^\mathrm{s}$  & 0.4 \\		
		\bottomrule
	\end{tabular}
\end{table}

%\begin{figure}[!h]
%\centering
%     \includegraphics[width=0.9\textwidth]{foilSchematic.eps}
%		\caption{The schematic of the cylinder-foil system.}
%		\label{Schematicbenchmark}
%\end{figure}

The boundary conditions for this FSI setup are identical to the benchmark case 
presented in \cite{turek2006proposal}. 
The no-slip Dirichlet condition is implemented on the surface of 
the cylinder wall, the flexible foil and on the top and the bottom surfaces. 
Of particular interest is the fluid-solid interface between 
the flexible foil and the fluid subdomain.
Traction-free condition is implemented on the outlet $\Gamma_{out}$. 
A parabolic velocity profile is specified at $\Gamma_{in}$:
\begin{equation}
u^\mathrm{f}(0,y)=1.5\bar{U}\frac{y(H-y)}{(\frac{H}{2})^2}=1.5\bar{U}\frac{4.0}{0.1681}y(0.41-y),
\label{eq:inlet}
\end{equation} 
where $u^\mathrm{f}$ is the $X$-component of the fluid velocity $\bubar^\mathrm{f} = (u^\mathrm{f},v^\mathrm{f},w^\mathrm{f})$, $\bar{U}$ denotes the mean inlet velocity and $H=4.1D$ is the height of the computational domain between $\Gamma_{top}$ and $\Gamma_{bottom}$.
%To characterize the fluid-structure physics of the flexible foil, 
%the three dimensionless parameters, namely the Reynolds number, 
%the bending rigidity $K_B$ and structure-to-fluid density ratio $\rho^\mathrm{r}$ 
%can be defined as:
%\begin{equation}
%	Re = \frac{\rho^\mathrm{f} \bar{U} D}{\mu^\mathrm{f}}, \; \; 
%        K_B = \frac{E w^3}{12(1-(\nu^\mathrm{s})^2) \rho^\mathrm{f} \bar{U}^2 L^3}, \; \; 
%        \rho^\mathrm{r} = \frac{\rho^\mathrm{s}}{\rho^\mathrm{f}},
%\end{equation}
%where $D$ denotes the diameter of the circular cylinder, 
%$w$ is the foil thickness, $L$ is the length of the foil,
%$E$ is the Young's Modulus, and
%$\nu^\mathrm{s}$ denotes the Poisson's ratio.

\begin{table}[!h]
	\centering
	\caption{Mesh convergence and validation of FSI-III  case. 
        The percentage differences are calculated by using M3 result as the reference}
			 \label{foilmesh}
	\begin{tabular}{cccccc}
		\toprule
		\multirow{2}{*}{Mesh} & Fluid &  Solid & \multirow{2}{*}{$A_{x,max}$} &  \multirow{2}{*}{$A_{y,max}$} &  \multirow{2}{*}{$f_{y}$} \\
		& elements & elements & & & \\
		\midrule 
		M1 & 7171  & 136  & 2.68 (9.39\%) & 32.94 (0.15\%) & 4.966 (8.71\%) \\
		M2 & 11770  & 544 & 2.55 (4.08\%) & 31.89 (3.04\%)& 5.267 (3.18\%) \\
		M3 & 25882  & 1280  & 2.45 & 32.89 & 5.44 \\
		Benchmark & 9216 & 1280 & 2.68 & 35.34 & 5.3 \\
		\bottomrule
	\end{tabular}
\end{table}

\begin{figure}[H]
	\centering
    \includegraphics[width=0.7\textwidth]{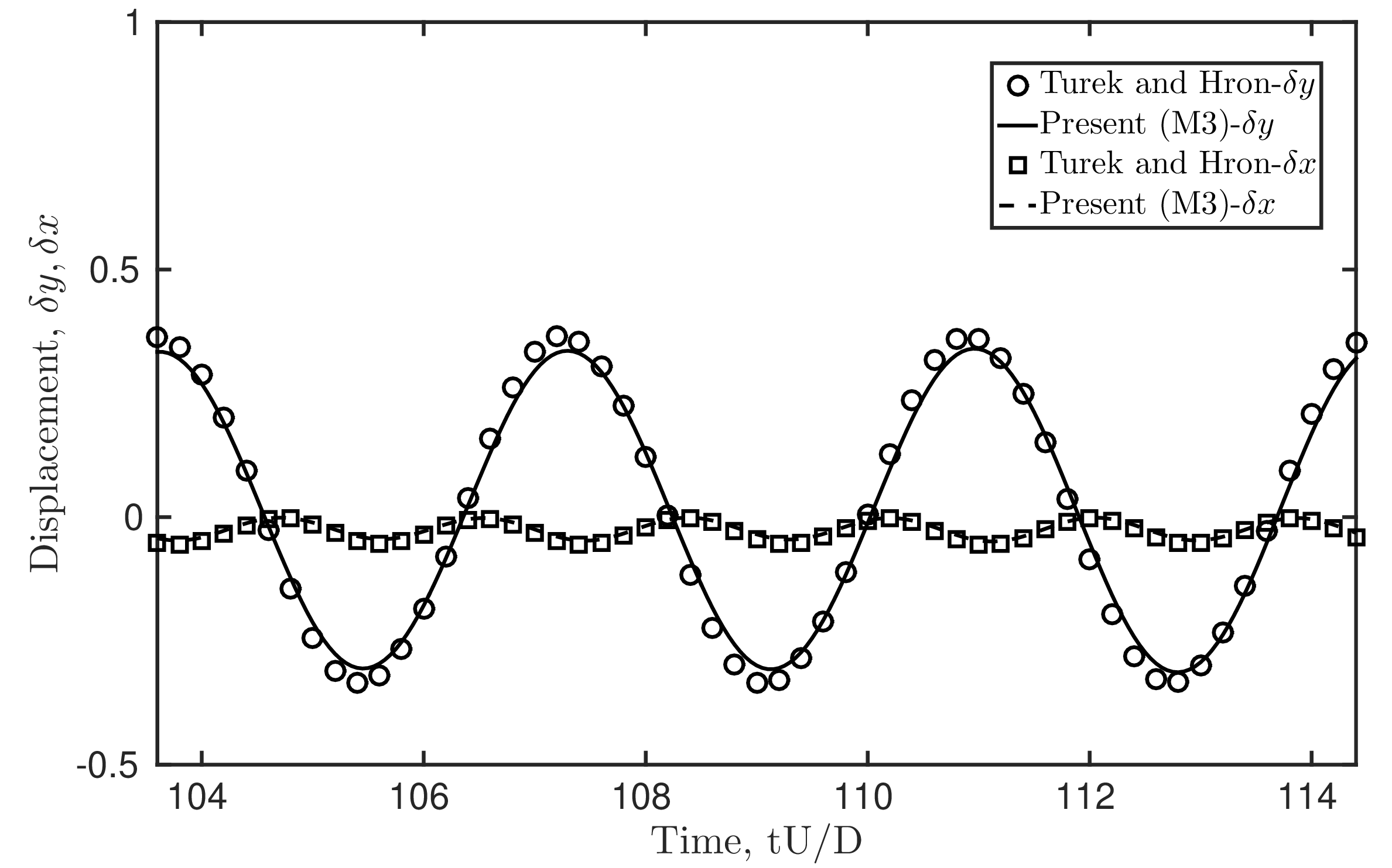}
		\caption{Comparisons of tip displacements between the present study using M3 mesh 
                and the benchmark data.}
		\label{turektipdisp}
\end{figure}

We decompose both the fluid and solid domains through finite element meshes. 
There is a boundary layer mesh surrounding the cylinder-foil 
and a triangular mesh outside the boundary layer region.
Mesh convergence study is conducted to ensure that sufficient mesh resolution is employed for both the fluid and solid subdomains. To eliminate the effect of the non-matching discretization, matching meshes are used to verify the fluid-structure 
coupling. Figure \ref{figFoilMesh} shows a typical mesh 
used for the verification study.  
%\begin{figure}[H]
%	\centering
%{\includegraphics[width=0.75\textwidth]{fsi3_matching_136.eps}} \; \; 
%	\caption{Cylinder-foil system matching mesh M1 dividing for numerical simulation.}
%	\label{fsi3_matching_136}
%\end{figure}
Three sets of matching meshes with increasing element number are selected 
for this study. Their characteristic responses are shown in Table \ref{foilmesh}. 
It is concluded that M3 has achieved sufficient convergence, 
therefore it is used as the reference case in our study. 
The computed values of maximum tip displacements ($A_{x,max}$ and $A_{y,max}$) 
and the transverse frequency $f_{y}$ 
are overall in good agreement with the benchmark solutions.
%
%Note that there is some discrepancy between the result of M3 and the benchmark 
%result. 
%
%This minor difference may be due to the difference in the hyperelastic model 
%used in the present simulation instead of 
%St. Venant-Kirchhoff model used in the benchmark computation. 
%
To further verify the accuracy of FSI solver for M3 mesh, 
the tip displacement history of the foil is plotted against the benchmark 
result in Fig. \ref{turektipdisp}. There is good agreement 
for both the displacement components between the present and the benchmark solutions.

\begin{table}[!h]
	\centering
	\caption{ Sensitivity and assessment of FSI results for varying mesh ratios $h_s/h_f$ }
			 \label{nonmatchingresult}
	\begin{tabular}{cccccc}
		\toprule
		$h_s/h_f$ & Fluid elements &  Solid elements & $A_{x,max}$ &  $A_{y,max}$ &  $f_{y}$\\
		\midrule
		4.0 & 25882  & 136  & 2.45 & 32.73 & 5.467 \\

		2.0 & 25882  & 544 & 2.47 & 32.97 & 5.455 \\

		1.0 & 25882  & 1280  & 2.45 & 32.89 & 5.468 \\
		
		0.5 & 25882  & 5120  & 2.46 & 32.95 & 5.455 \\
				
		0.25 & 25882  & 20480  & 2.44 & 32.98 & 5.442 \\
		\bottomrule
	\end{tabular}
\end{table}

\begin{figure}[!h]
	\centering
	\subfloat[][X-Y view]
{\includegraphics[width=0.75\textwidth]{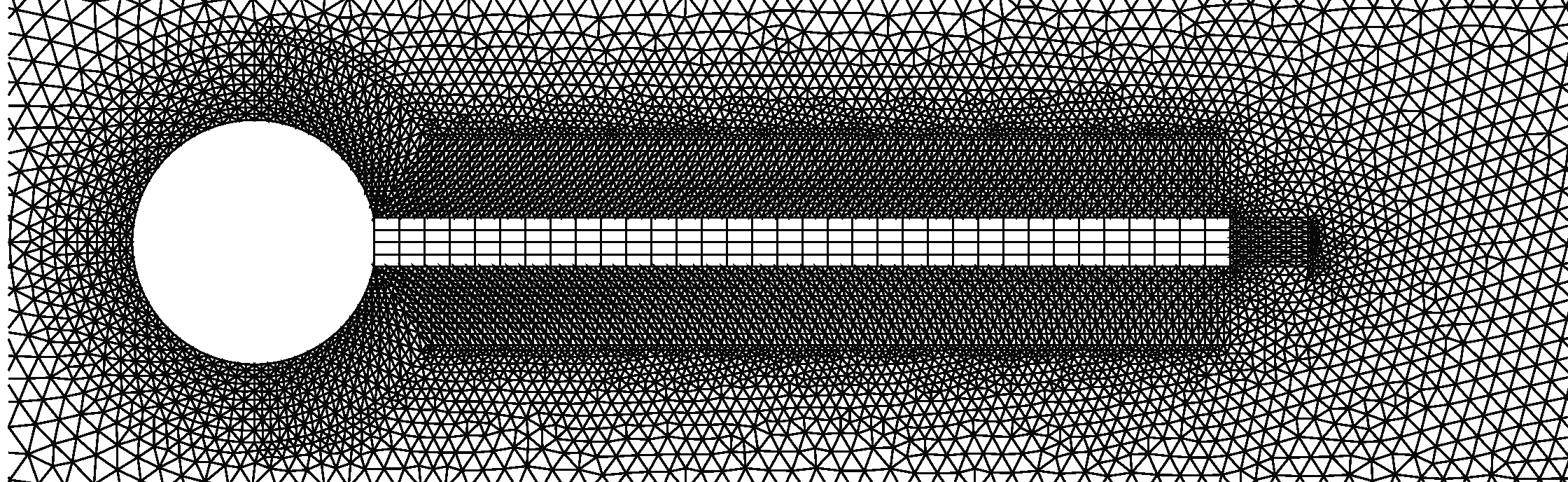}} \; \; 
	\subfloat[][Y-Z view]{\includegraphics[width=0.75\textwidth]{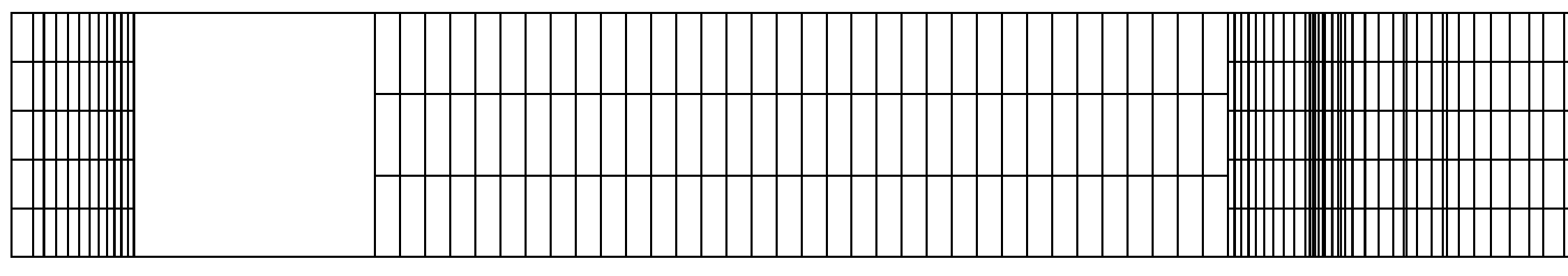}} \; \;
	\caption{A representative non-matching mesh configuration for the cylinder-foil system.}
	\label{fsi3_nonmatching_136}
\end{figure}

Next, we proceed to quantify the accuracy of the FSI simulation when non-matching meshes 
with different mesh ratios are employed. 
A typical mesh used for this study is shown in Fig. \ref{fsi3_nonmatching_136}.
A series of numerical experiments is conducted by considering the mesh ratio ranging from $0.25$ to $4.0$ 
to capture both the situations where the load is transferred from 
the coarse mesh to the fine mesh and vice-versa. 
The characteristic response data of tip displacements and frequency for each mesh ratio $h_s/h_f$ 
are summarized in Table \ref{nonmatchingresult}.

%\begin{figure}[H]
%	\centering
%	\subfloat[][drag coefficient Cd]{\includegraphics[width=0.45\textwidth]{f1280ratioDispX.eps}} \; \; 
%	\subfloat[][rms lift coefficient Cl]{\includegraphics[width=0.45\textwidth]{f1280ratioDispY.eps}}
%	\caption{force coefficients.}
%	\label{fsi3nonmatchingdisp}
%\end{figure}

\begin{figure}[h!]
	\centering
	\subfloat[][]{\includegraphics[width=0.3\textwidth]{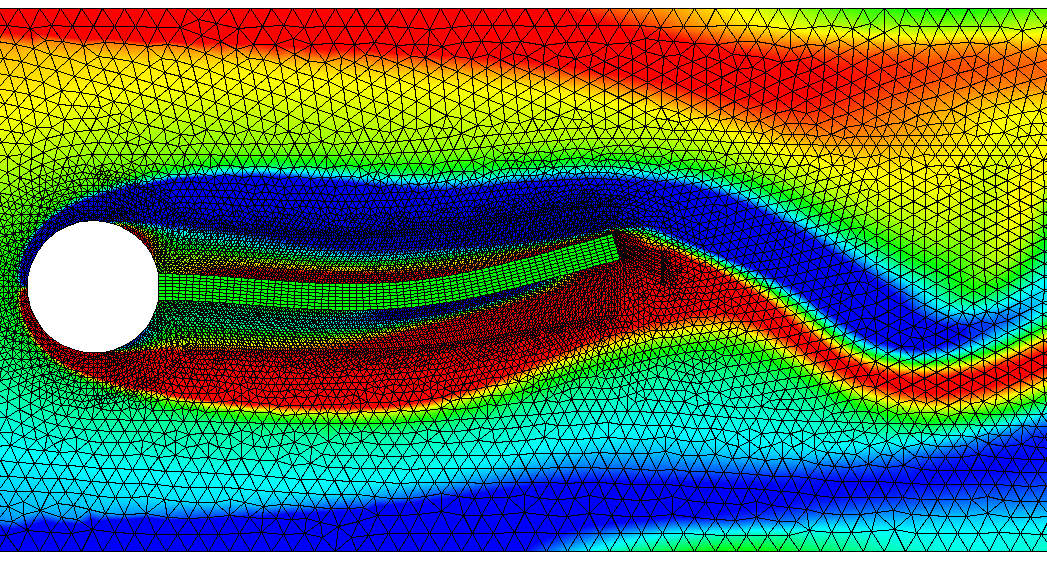}} \;
	\subfloat[][]{\includegraphics[width=0.3\textwidth]{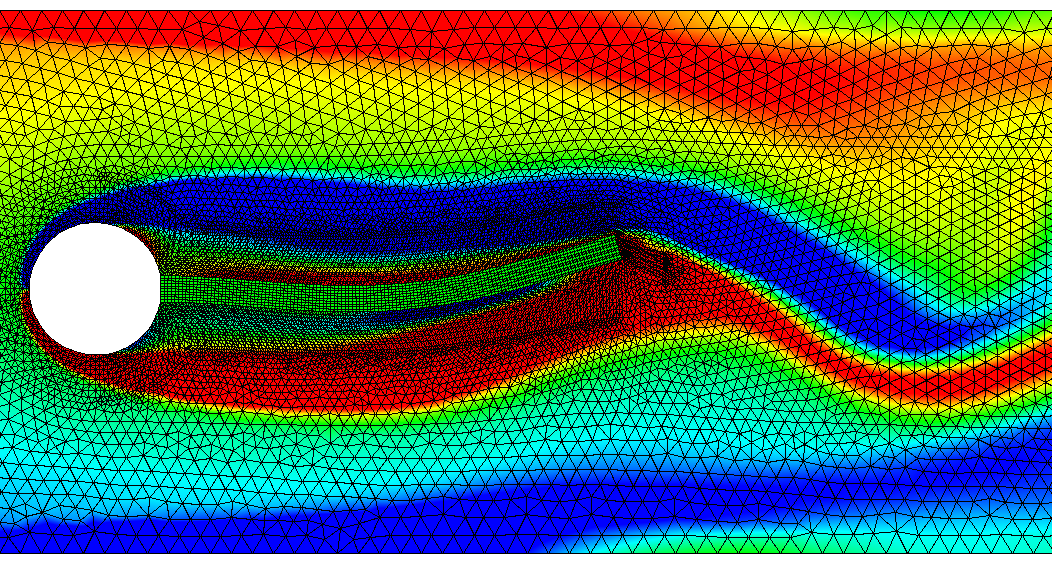}} \;
	\subfloat[][]{\includegraphics[width=0.3\textwidth]{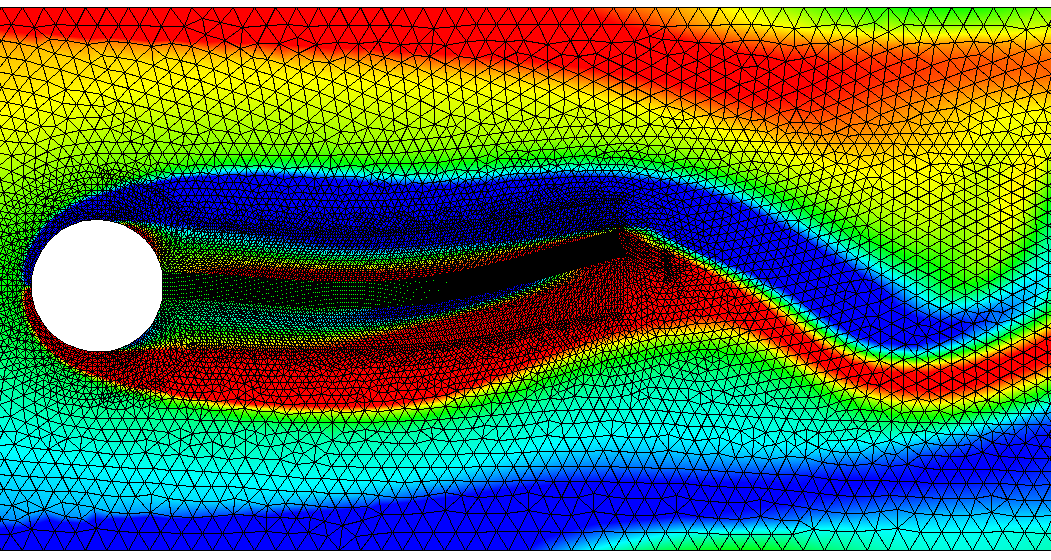}} \\
	\subfloat{\includegraphics[width=0.5\textwidth]{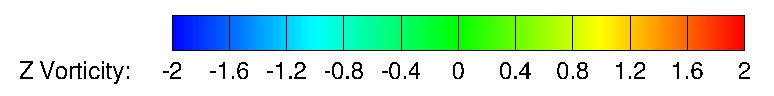}}
	\caption{Instantaneous $Z$-vorticity contours and meshes for representative mesh ratio: (a) fine to coarse mesh, $h_s/h_f = 2.0$, (b) matching mesh, $h_s/h_f = 1.0$, (c) coarse to fine mesh, $h_s/h_f = 0.5$. 
        }
	\label{figFoilZvorMesh}
\end{figure}

It is evident that the characteristic response values are close to each other for the range 
of mesh ratios $h_s/h_f$. Comparison of instantaneous $Z$-vorticity contours among 
the representative mesh ratios ($h_s/h_f = 0.5, 1.0, 2.0$) is illustrated 
in Fig. \ref{figFoilZvorMesh}. These flow contours are plotted at the instant 
where the tip of the foil reaches its maximum displacement.
For all practical purposes, we can see that the flow patterns are very similar for different mesh ratios, 
thus indicating the qualitative accuracy of the 3D common-refinement.
Next we demonstrate the accuracy and stability of the proposed FSI formulation 
to large-scale 3D simulation of long flexible offshore riser.

%%%%%%%%%%%%%%%%%%%%%%%%%%%%%%%%%%%%%%%%%%%%%%%%%%%%%%%%%%%%%%%%%%%%%%%%%%%%%%%%%%%%%%%%%%%%
\section{Application to offshore riser VIV}\label{sec:riser}
\begin{figure}[h!]
	\centering
	\subfloat[][]{\includegraphics[width=0.55\textwidth]{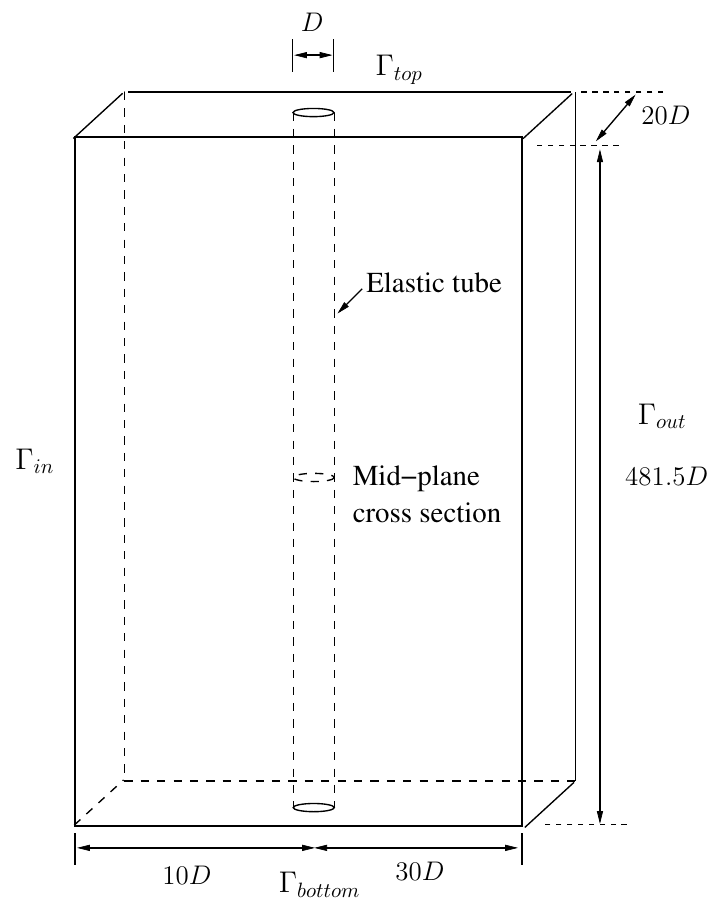}} \qquad
	\subfloat[][]{\includegraphics[width=0.38\textwidth]{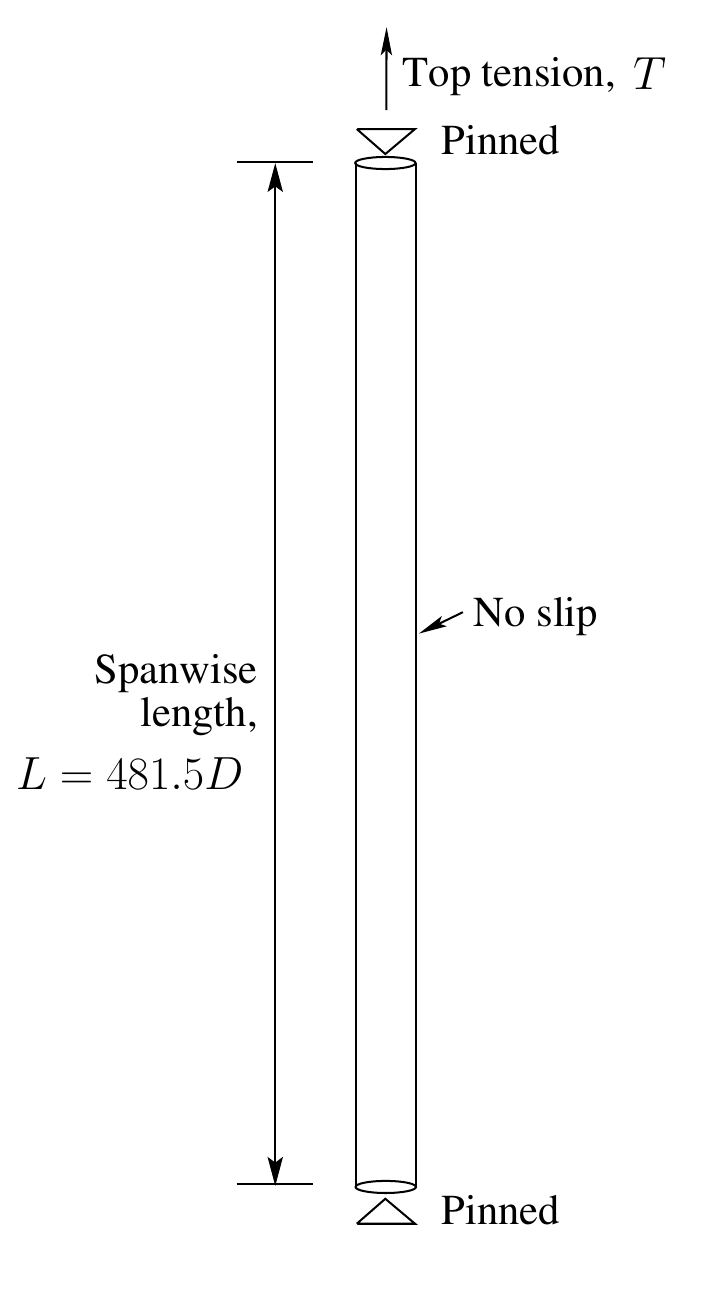}}
	\caption{Flow past a flexible offshore riser: (a) schematic of computational setup, 
        (b) boundary conditions applied on the pinned-pinned tensioned riser.}
	\label{figLongRiserSchematic}
\end{figure}

The prediction of VIV of an offshore riser is crucial to prevent operational failure in complex ocean 
environment. In this section, we demonstrate that our fluid-solid coupled FEM solver with 
the 3D common-refinement scheme is capable of predicting the dynamics of a riser with reasonable accuracy. 
To achieve this, we simulate a riser under uniform flow conditions which was carried out as a part of 
an experimental campaign in \cite{riserExperiment}. A schematic diagram of the riser setup is shown 
in Fig. \ref{figLongRiserSchematic}. Consider the diameter of the circular riser to be $D$. 
The inlet ($\Gamma_{in}$) and outlet ($\Gamma_{out}$) boundaries are at a distance of 10$D$ and 30$D$ 
from the center of the cylindrical riser. The side boundaries are equidistant from the center of the riser 
at 10$D$ corresponding to 5\% blockage ratio. The riser spans in $Z$-direction, with its spanwise length, $L$ equal to 481.5$D$. Let the components of the fluid velocity be given as $\bubar^\mathrm{f} = (u^\mathrm{f},v^\mathrm{f},w^\mathrm{f})$. A freestream velocity $u^\mathrm{f}=U$ along the $X$-axis is imposed at the 
inlet boundary $\Gamma_{in}$. The top and bottom boundaries, $\Gamma_{top}$ and $\Gamma_{bottom}$ 
have slip boundary condition, where $\pd{u^\mathrm{f}}{y} = 0$ and $v^\mathrm{f}=0$. 
The outlet $\Gamma_{out}$ has a traction-free boundary condition, where $\sigma_{xx}=\sigma_{yx}=\sigma_{zx} = 0$. Pinned-pinned boundary condition is implemented on both the ends of the riser, with a tension, 
$T$ applied at the top of the riser, while no-slip condition is applied on the riser surface.

\begin{figure}[h!]
	\centering
	\includegraphics[width=0.95\textwidth]{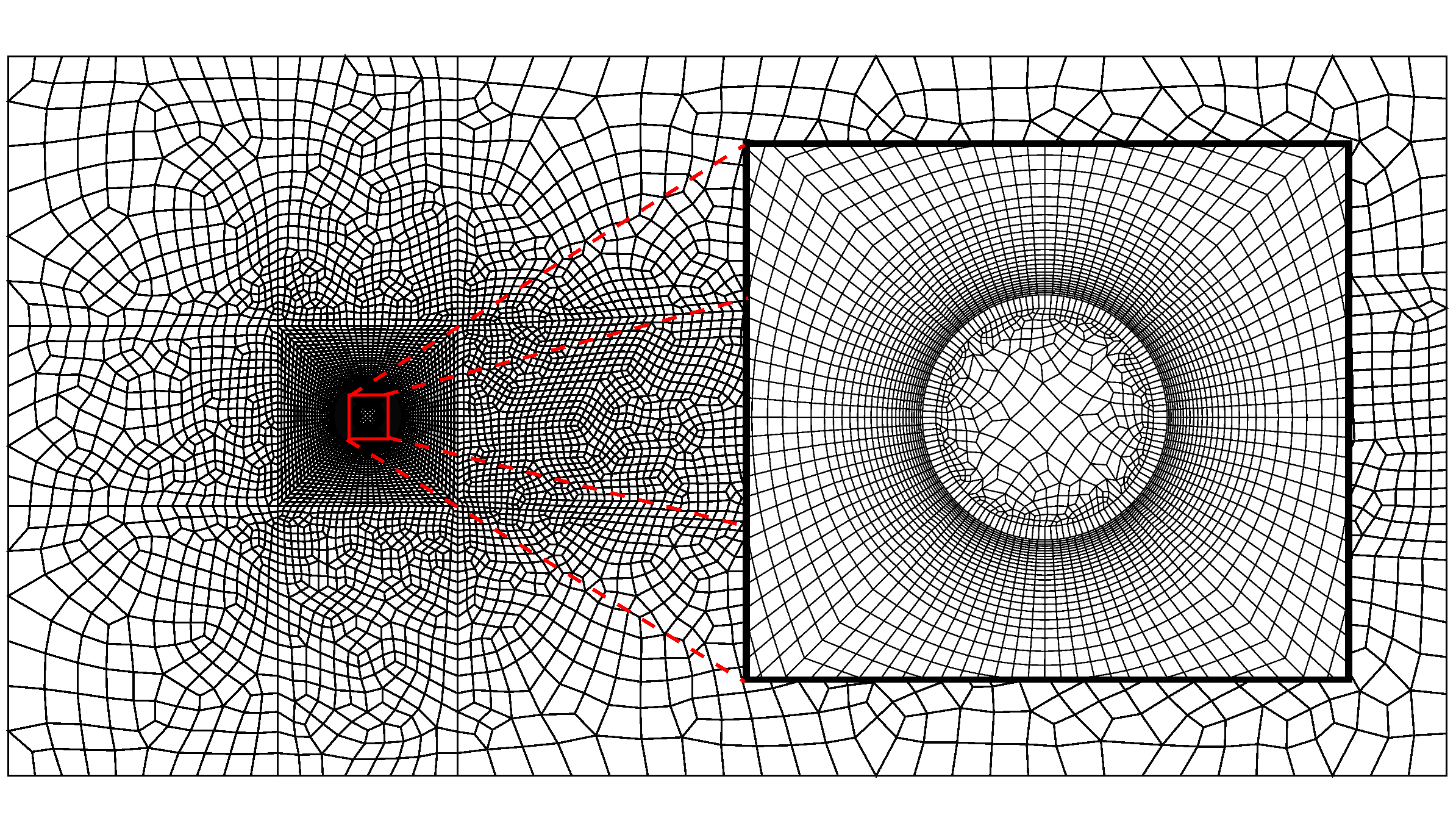}
	\caption{A long riser under uniform flow: two-dimensional layer of the unstructured non-matching computational mesh. The inset shows the magnified view of the non-matching fluid-structure interface. The mesh is extruded in the third-dimension while maintaining a non-matching spanwise mesh.}
	\label{figLongRiserMesh}
\end{figure}

A representative discretization of the fluid and structural mesh is shown in Fig. \ref{figLongRiserMesh}. 
The number of divisions on the circumference of the cross-section of the riser on the fluid and solid meshes 
is 96 and 120 respectively. A boundary layer is maintained along the riser such that $y^+<1$ in
the wall-normal direction. The fluid domain is discretized into $9 \times 10^5$ nodes with $1.2 \times 10^6$ unstructured hexahedral elements, and the riser is discretized into $8.7 \times 10^4$ nodes with $9.7 \times 10^4$ hexahedral elements. The non-dimensional time step is selected as $\Delta tU/D = 0.1$. 
The dimensionless parameters used in the simulation are:
\begin{align}
	Re &= \frac{\rho^\mathrm{f} U D}{\mu^\mathrm{f}} = 4000, \\
	\frac{EI}{\rho^\mathrm{f} U^2 D^4} &= 2.1158 \times 10^7, \\
	\frac{T}{\rho^\mathrm{f} U^2 D^2} &= 5.10625 \times 10^4, \\
	m^* &= \frac{m^\mathrm{s}}{\frac{\pi}{4} D^2 L \rho^\mathrm{f}} = 2.23, 
\end{align}
where $I$ is the second moment of area of its cross section, $m^\mathrm{s}$ is the mass of the riser. 
The result of the simulation is discussed in the following sections.

\subsection{Response characteristics}
\begin{figure}[h!]
	\centering
	\includegraphics[width=0.7\textwidth]{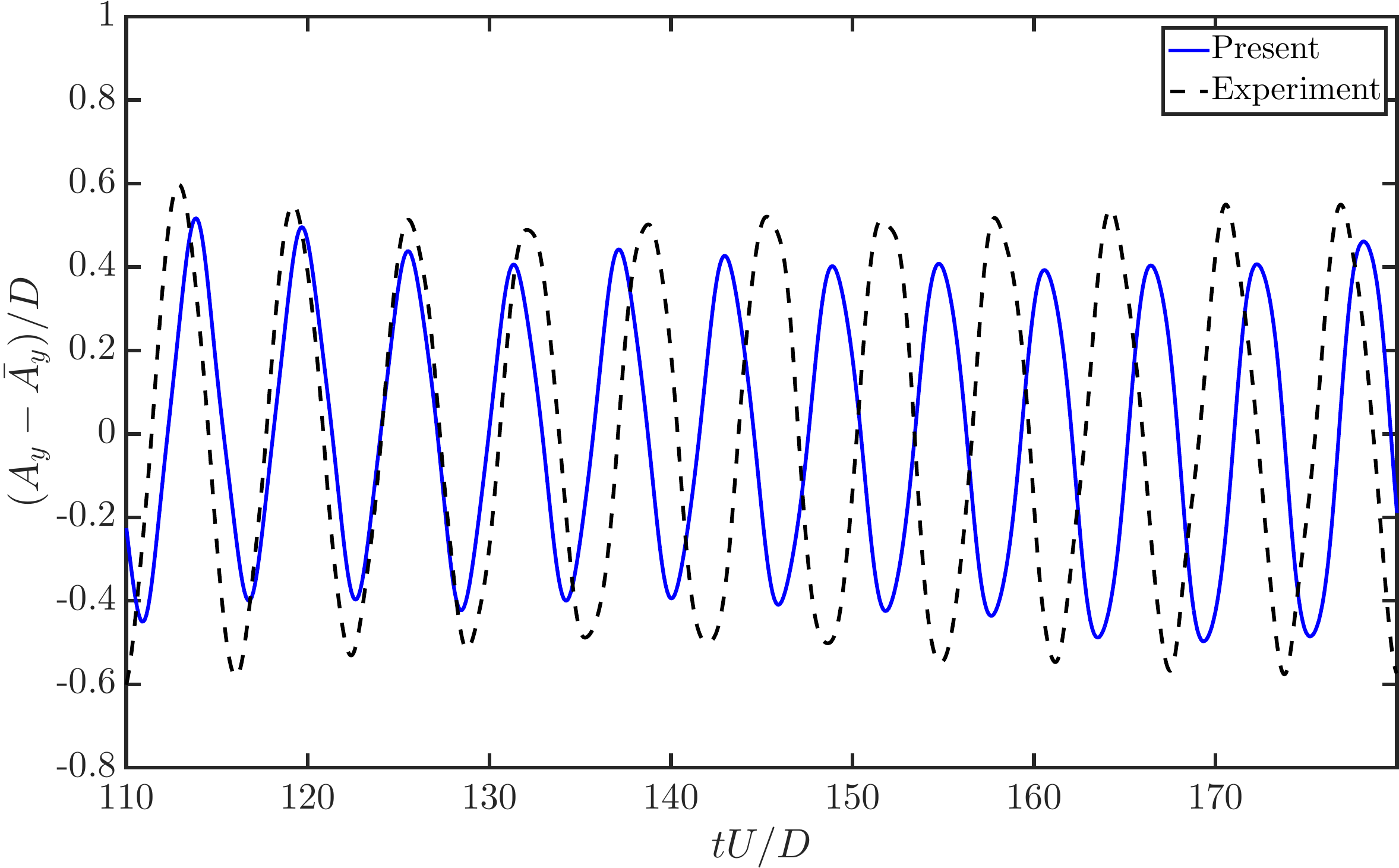}
	\caption{Riser response under uniform current flow: time history of the  cross-flow displacement at $z/L = 0.55$.}
	\label{figYHistory}
\end{figure}

\begin{figure}[h!]
	\centering
	\subfloat[][]{\includegraphics[width=0.50\textwidth]{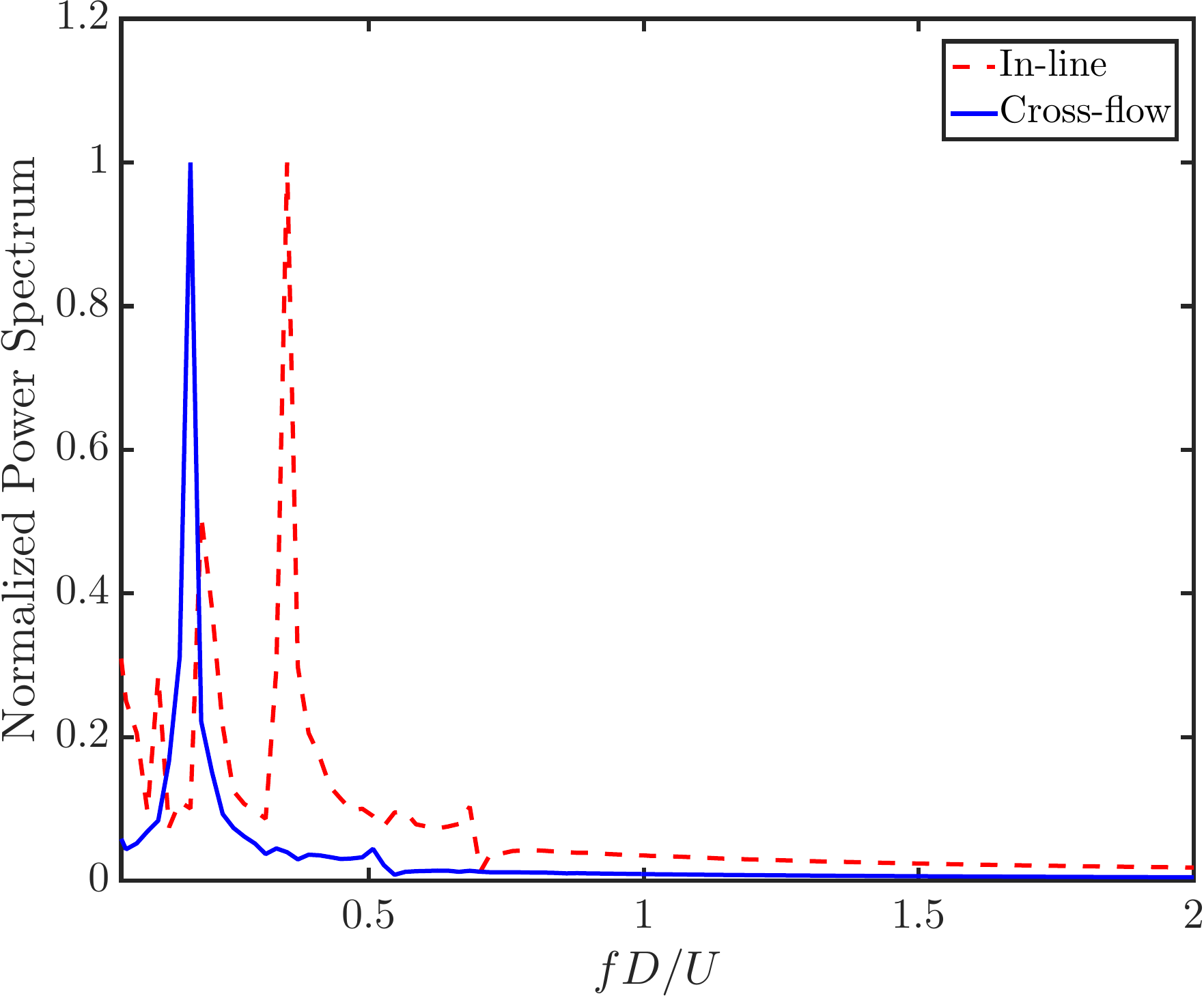} \label{figFreq}}
	\subfloat[][]{\includegraphics[width=0.49\textwidth]{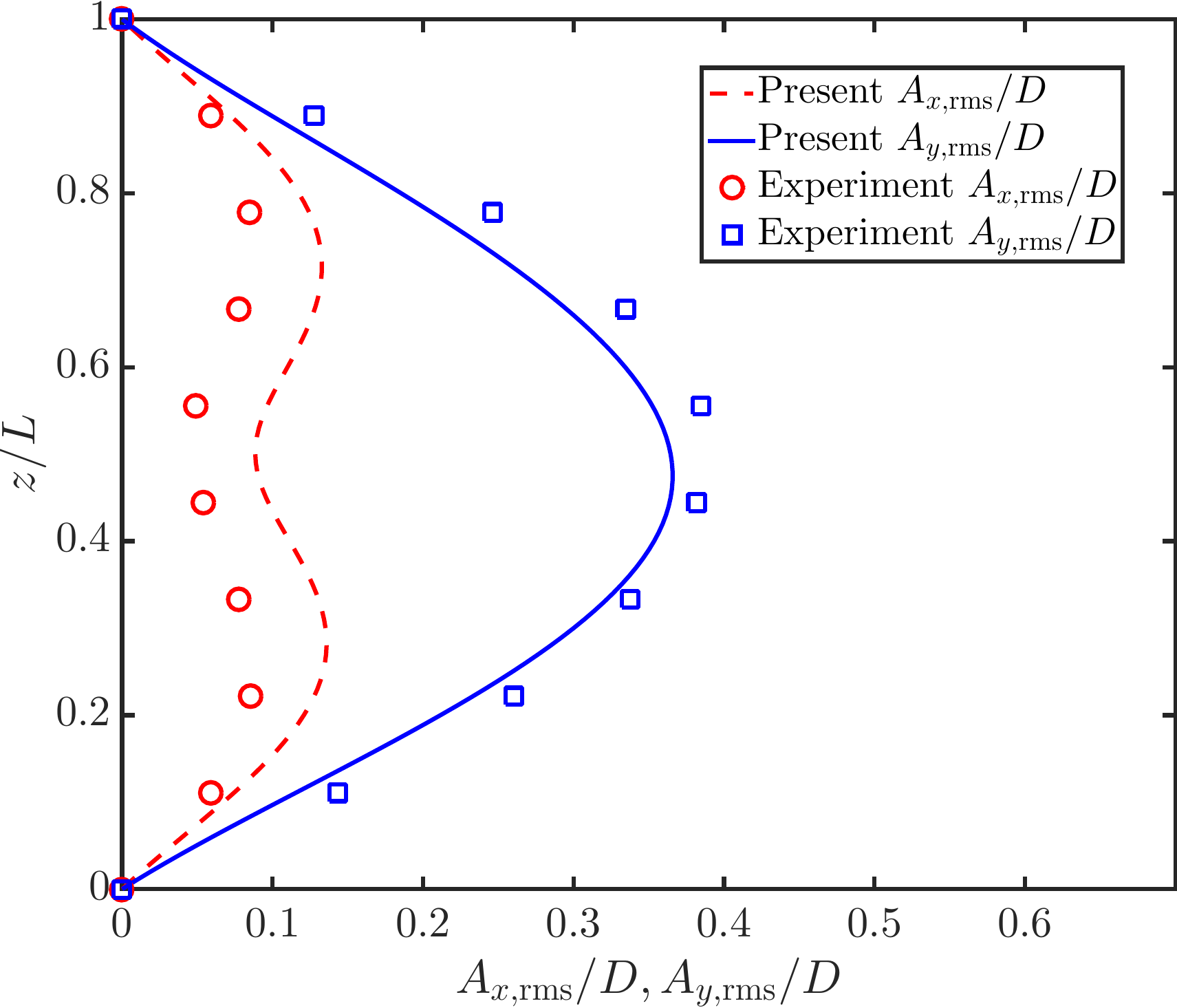} \label{figYrms}} 
	\caption{Riser response under uniform current flow at $Re=4000$: 
        (a) power spectrum of the in-line and the cross-flow amplitudes along the riser, 
        (b) comparison of the root mean square values of the in-line and cross-flow displacements 
        along the riser with that of the experiment.}
	\label{figRiserAmp}
\end{figure}

The response along the riser is analyzed and compared to that of the experiment in this section. 
Figure \ref{figYHistory} shows a comparison of the time history of the cross-flow displacement at the position $z/L = 0.55$ between the present simulation and the experiment. We observe a multi-frequency response which is reflected by the spectral analysis carried out for the identical response in Fig. \ref{figFreq}. Furthermore, we observe that the in-line response frequency ($fD/U = 0.3516$) is twice 
that of the cross-flow frequency ($fD/U = 0.1758$).
We also compare the root mean square values of the riser response with that of the experiment. Consider $A_x$ and $A_y$ as the displacement amplitudes at a point along the riser in the in-line and cross-flow directions respectively. Let the temporal mean of the in-line and cross-flow amplitudes at the location be denoted by $\overline{A_x}$ and $\overline{A_y}$ respectively. Then, the root mean square values are calculated as
\begin{align}
	A_{x,\mathrm{rms}} = \sqrt{\frac{1}{N}\displaystyle\sum^{N}_{i=1} (A_{x,i} - \overline{A_x})^2},\\
	A_{y,\mathrm{rms}} = \sqrt{\frac{1}{N}\displaystyle\sum^{N}_{i=1} (A_{y,i} - \overline{A_y})^2},
\end{align}
where $N$ represents the number of samples collected in time. The rms values are plotted in Fig. \ref{figYrms}.
Although it shows a good agreement for the cross-flow amplitude, some over-prediction of the in-line amplitude is observed. 
% The difference in the maximum in-line rms value between the numerical and experimental result is $< --\%$ with respect to the total maximum rms value $A$ given by $A=\sqrt{(A_{x,\mathrm{rms}}^2+A_{y,\mathrm{rms}}^2)}$. 
This difference may be due to the sensitivity in the measurement of the in-line amplitude and boundary layer characteristics along the flexible riser. 

\begin{figure}[h!]
	\centering
	\subfloat[][]{\includegraphics[width=0.49\textwidth]{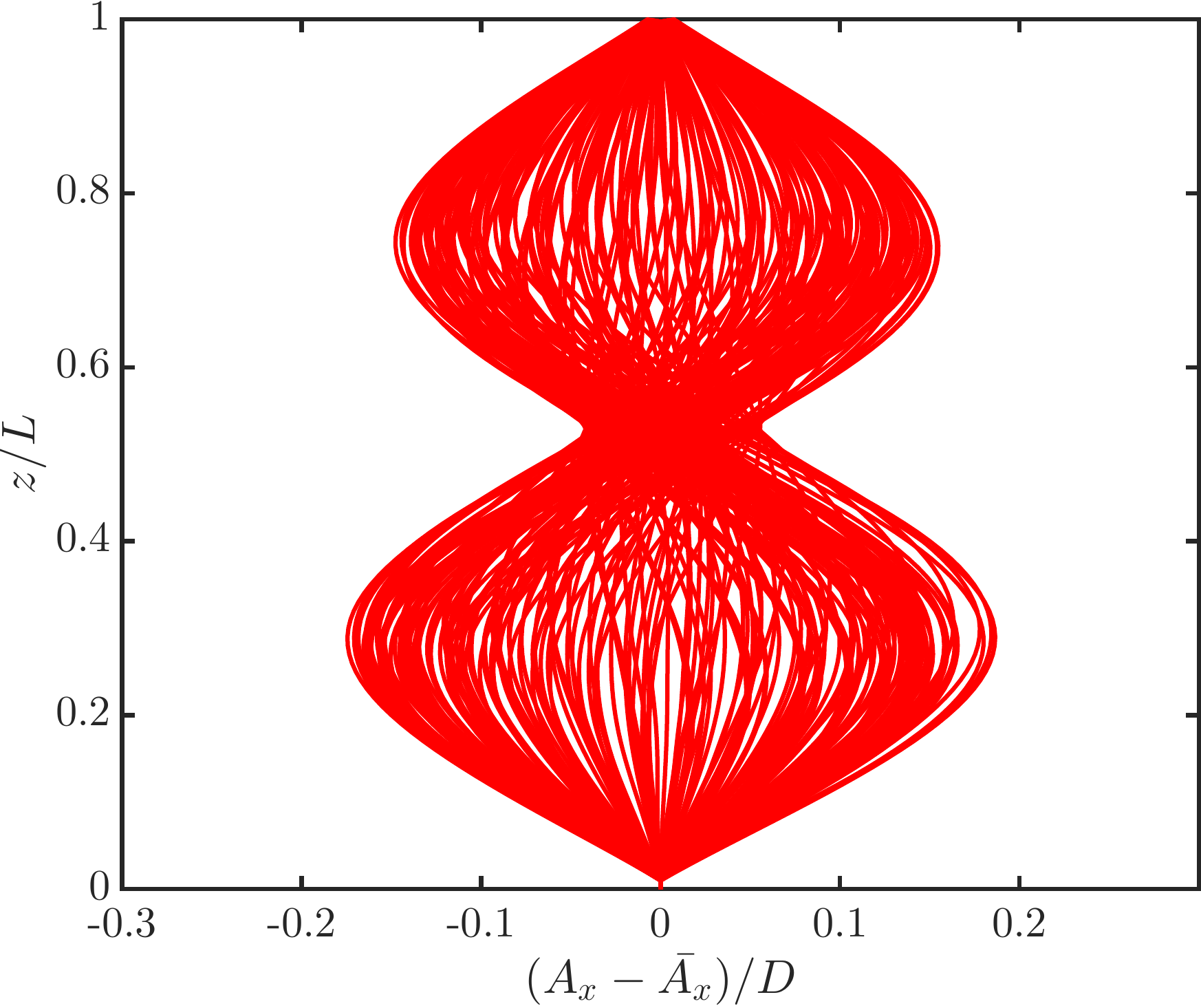} \label{figxshape}} 
	\subfloat[][]{\includegraphics[width=0.49\textwidth]{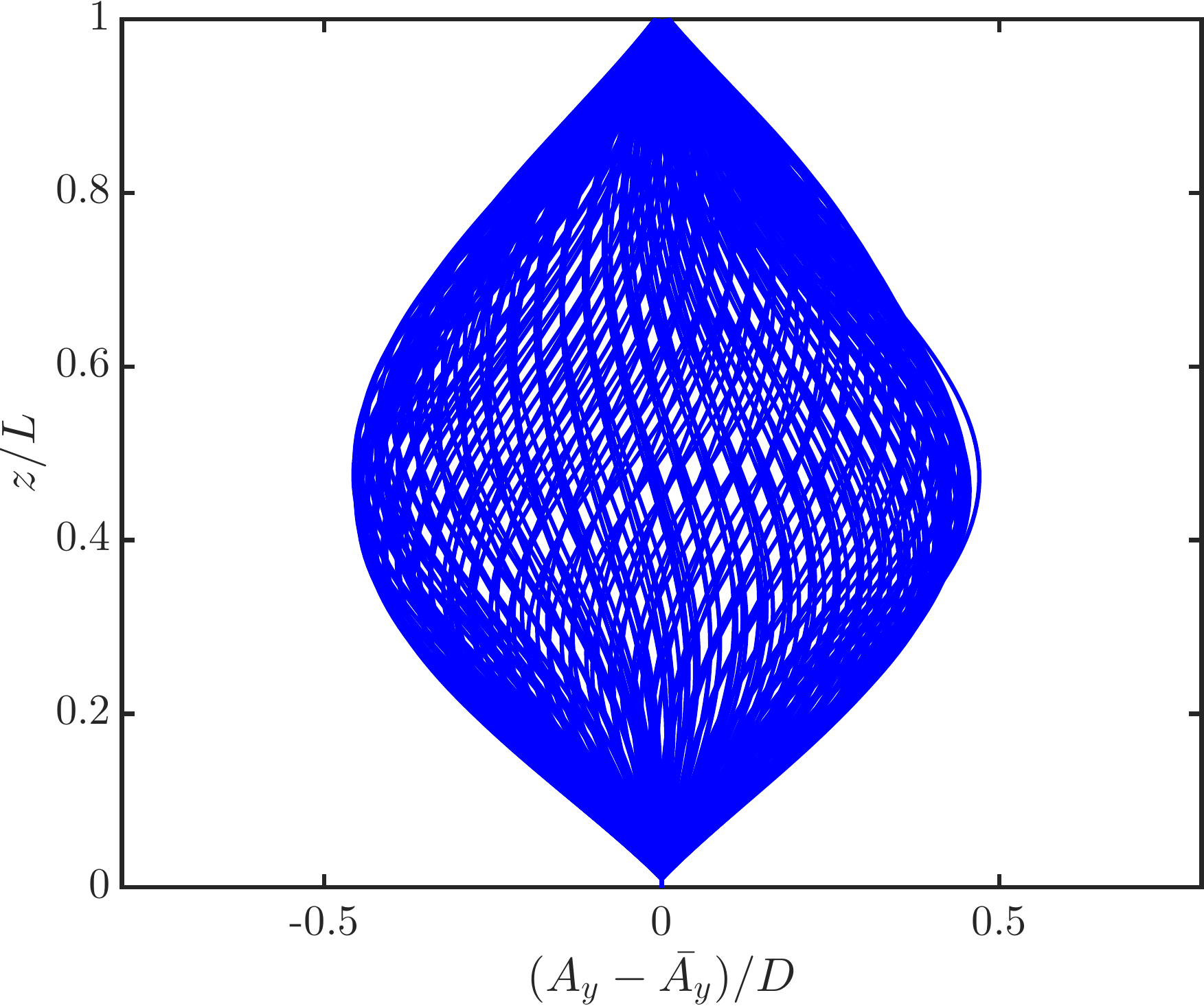} \label{figyshape}} \\ 
	\caption{Riser response envelope under uniform current flow: (a) in-line and (b) cross-flow directions. The riser is vibrating in the fundamental mode for the cross-flow, and the second mode for the in-line directions.}
	\label{figRiserShape}
\end{figure}

\begin{figure}[h!]
	\centering
	\subfloat[][]{\includegraphics[trim={2cm 0 2cm 0cm},clip,width=0.50\textwidth]{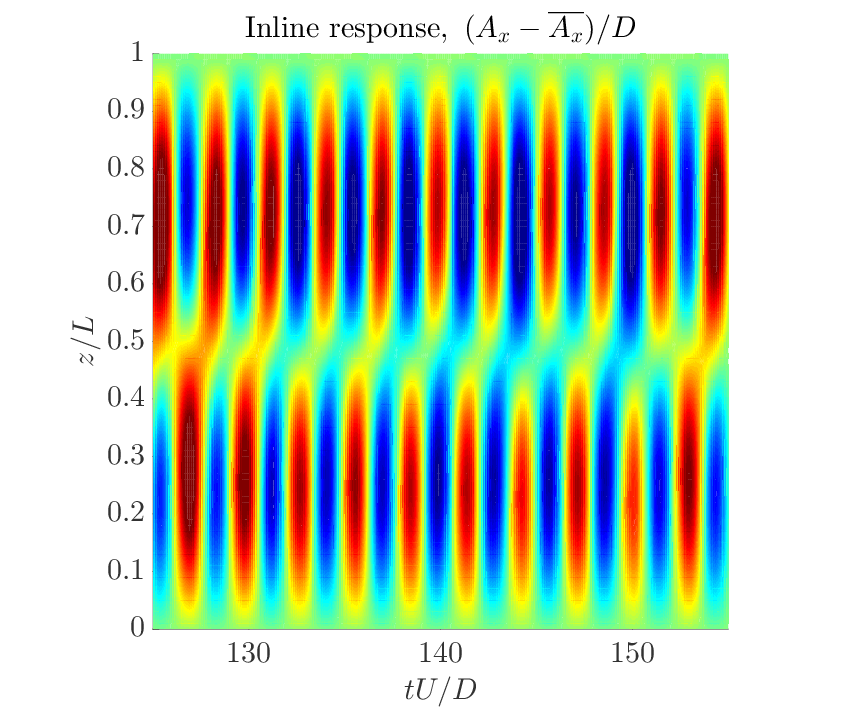} \label{figXwaves}} 
	\subfloat[][]{\includegraphics[trim={2cm 0 2cm 0cm},clip,width=0.50\textwidth]{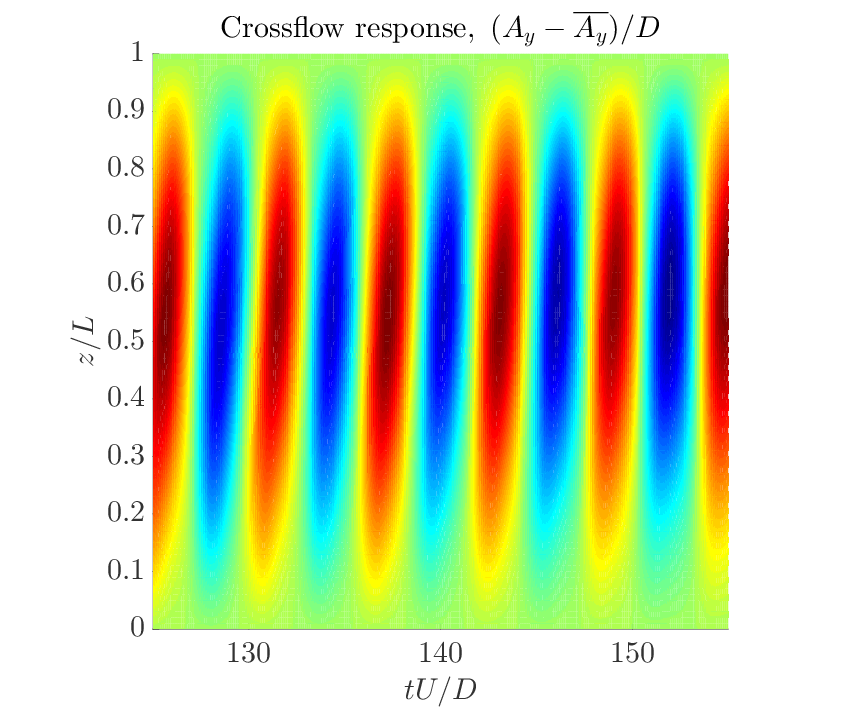} \label{figYwaves}}
	\caption{Standing wave riser response under uniform current flow: (a) in-line; (b) cross-flow.}
	\label{figRiserWaves}
\end{figure}

The response envelope of the riser is depicted in Fig. \ref{figRiserShape}. We infer that the riser vibrates in a dominant second mode in the in-line direction and the first mode in the cross-flow direction which corroborates our observation of dual resonance from the spectral study. The riser response along the span of the riser with time is also shown in Fig. \ref{figRiserWaves}. We observe a standing wave-like pattern in the response. 

\subsection{Vortex patterns}
\begin{figure}[h!]
	\centering
	\subfloat[][]{\includegraphics[width=0.49\textwidth]{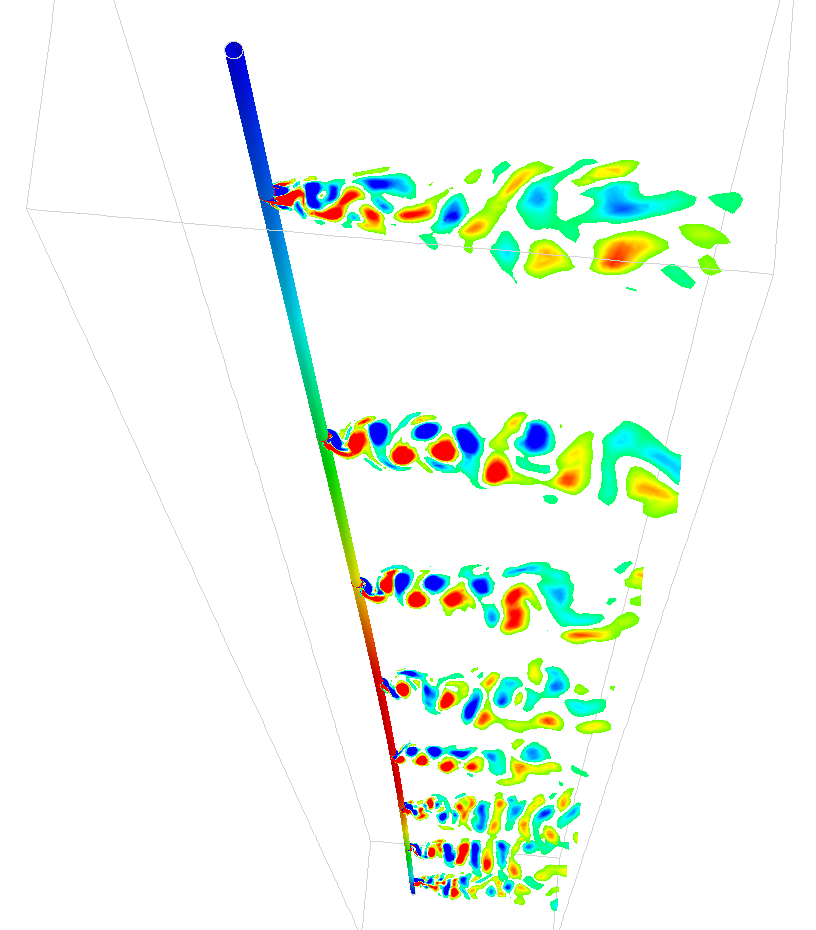} \label{figDispVor}} 
	\subfloat[][]{\includegraphics[width=0.49\textwidth]{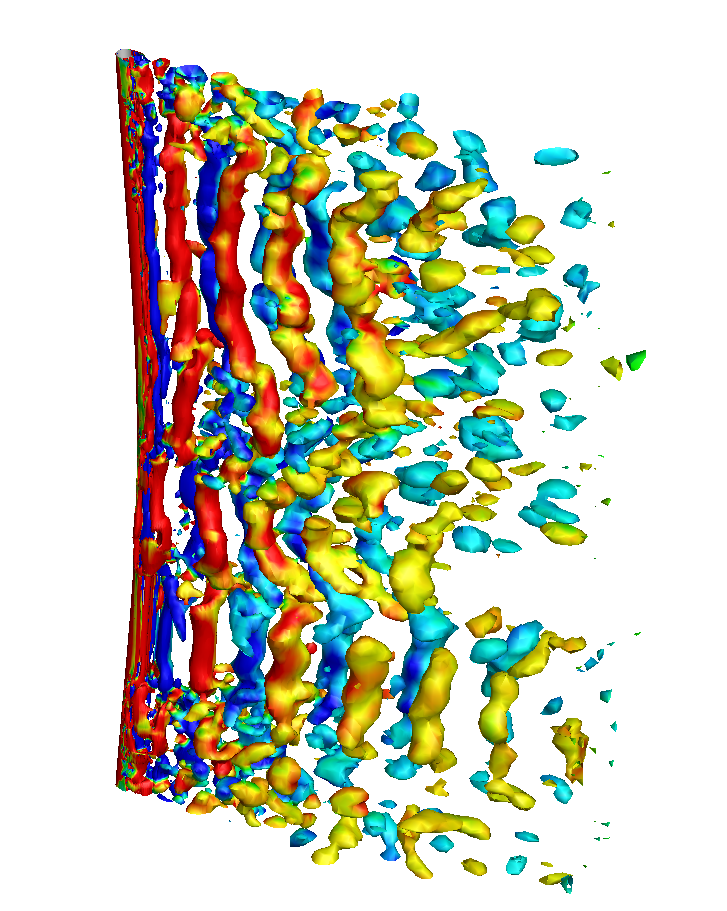} \label{figQcrit}} \\
	\subfloat{\includegraphics[width=0.55\textwidth]{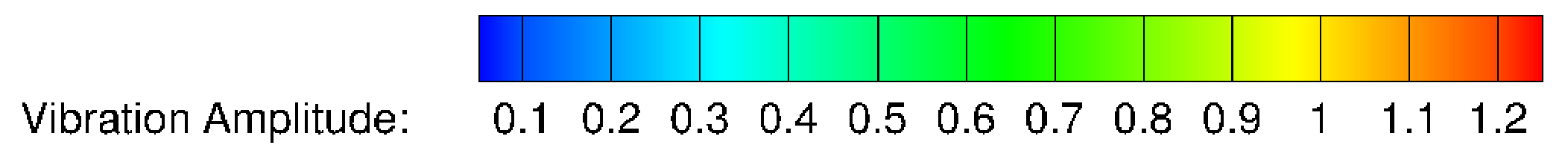}}
	\subfloat{\includegraphics[width=0.35\textwidth]{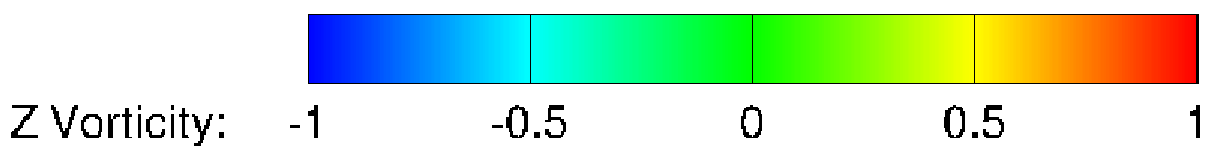}}  
	\caption{(a) Vibration amplitude (surface of riser) and $Z$-vorticity contours (cross section) at different positions along the riser; (b) Instantaneous isosurfaces of Q ($-\frac{1}{2} \pd{u_i}{x_j} \pd{u_j}{x_i}$) at $Q=0.1$, colored by $Z$-vorticity.}
	\label{figRiserVortex}
\end{figure}

In this section, we give some insight about the flow visualization and vortex patterns observed along the riser.
The $Z$-vorticity contours at locations $z/L \in [0.11,0.88]$ along with the vibration amplitude along the riser is displayed in Fig. \ref{figDispVor}. We observe the 2S mode of vortex shedding in most of the locations along the riser while more complex shedding patterns are observed near the location with large amplitude. A 2P vortex mode is also observed in some locations. Further analysis is required to study the vortex dynamics in detail. The isosurfaces of Q-criterion colored by the $Z$-vorticity are shown in Fig. \ref{figQcrit}. We observe more intense vortical structures near the location of large response amplitude. A more detailed analysis of the response amplitude and its relation to the trajectories along the riser can be found in \cite{vaibhav_CAF}.
This demonstration concludes that the present solver with common-refinement scheme is able to capture the physics as well as the response of the flexible riser reasonably well for non-matching unstructured meshes.

\section{Conclusions}\label{sec:conclusion}
In this paper, a 3D conservative data transfer scheme based on the common-refinement method 
is presented for the non-matching interface between partitioned finite-element subdomains of 
incompressible flow and nonlinear hyperelastic structure. 
Owing to the weighted residual formulation in three dimensions, the common-refinement method 
has a good conservative and accurate properties for fluid-structure interactions.
The current implementation can be used for a wide range of FSI applications
while retaining the modularity of both physical systems and the flexibility of non-matching meshes.
We have shown that the 3D common-refinement scheme provides an accurate coupling by minimizing 
the error in the transfer of data in the $L_2$ norm across the overlay surface between the fluid and the structure meshes.  
Through a systematic error analysis, the common-refinement method was shown to be consistent 
in terms of overall accuracy for three-dimensional data transfer. 
Second-order spatial accuracy of the common-refinement scheme has been confirmed 
for different mesh ratios.
The common-refinement method was then combined with the nonlinear iterative force correction procedure 
to solve FSI problems with strong added-mass effects associated with low structure-to-fluid density ratio.
The overall implementation was verified against the reference benchmark data for the cylinder-foil system.
Accuracy and stability of the combined common-refinement and NIFC has been demonstrated for the cylinder-foil 
system for a range of mesh ratios along the fluid-solid interface.
We demonstrated the applicability of the combined formulation to large 3D simulations of offshore 
riser problem with non-matching meshes.  The combined common-refinement and NIFC scheme 
provided a stable solution for the 3D flexible riser in a turbulent flow with strong inertial coupling 
of the surrounding wake flow. The cross-flow amplitude of flexible riser agreed reasonably 
well with the measurement data.  
To improve the efficiency of coupled FSI framework for high-gradient physics, 
it will be worth developing adaptive 3D non-matching meshes across the fluid-structure interface.

\section*{Acknowledgements}
The authors would like to thank the Singapore Maritime Institute and 
the National Research Foundation for the financial support.
The first author appreciates the help from  Prof. Allan Bower 
on hyperelastic theory and implementation.

\bibliographystyle{elsarticle-num}
\bibliography{refs}

\begin{thebibliography}{10}
\expandafter\ifx\csname url\endcsname\relax
  \def\url#1{\texttt{#1}}\fi
\expandafter\ifx\csname urlprefix\endcsname\relax\def\urlprefix{URL }\fi
\expandafter\ifx\csname href\endcsname\relax
  \def\href#1#2{#2} \def\path#1{#1}\fi

\bibitem{lee2005non}
S.-C. Lee, M.~N. Vouvakis, J.-F. Lee, A non-overlapping domain decomposition
  method with non-matching grids for modeling large finite antenna arrays, J.
  Comput. Phys. 203~(1) (2005) 1--21.

\bibitem{peng2010non}
Z.~Peng, J.-F. Lee, Non-conformal domain decomposition method with second-order
  transmission conditions for time-harmonic electromagnetics, J. Comput. Phys.
  229~(16) (2010) 5615--5629.

\bibitem{hueber2009thermo}
S.~H{\"u}eber, B.~Wohlmuth, Thermo-mechanical contact problems on non-matching
  meshes, Comp. Meth. Appl. Mech. Engrg. 198~(15) (2009) 1338--1350.

\bibitem{bathe}
N.~El-Abbasi, K.-J. Bathe, Stability and patch test performance of contact
  discretizations and a new solution algorithm, Comput. Struct. 79 (2001)
  1473--1486.

\bibitem{wohlmuth}
B.~Flemisch, M.~A. Puso, B.~I. Wohlmuth, A new dual mortar method for curved
  interfaces: 2{D} elasticity, Int. J. Numer. Meth. Engng. 63 (2005) 813--832.

\bibitem{aerothermo}
I.~Lee, J.~H. Roh, I.~K. Oh, Aerothermoelastic phenomena of aerospace and
  composite structures, Journal of Thermal Stresses 26 (2003) 526--546.

\bibitem{jaiman_jcp}
R.~Jaiman, X.~Jiao, X.~Geubelle, E.~Loth, Conservative load transfer along
  curved fluid-solid interface with nonmatching meshes, J. Comput. Phys. 218
  (2006) 372--397.

\bibitem{jaiman_omae2}
R.~Jaiman, F.~Shakib, O.~Oakley, Y.~Constantinides, Fully coupled
  fluid-structure interaction for offshore applications, in: ASME Offshore
  Mechanics and Arctic Engineering OMAE09-79804 CP, 2009.

\bibitem{law2017wake}
Y.~Z. Law, R.~K. Jaiman, Wake stabilization mechanism of low-drag suppression
  devices for vortex-induced vibration, Journal of Fluids and Structures 70
  (2017) 428--449.

\bibitem{blevins1990flow}
R.~D. Blevins, Flow-induced vibration, Van Nostrand Reinhold Co., Inc., New
  York, 1990.

\bibitem{blom}
F.~J. Blom, A monolithical fluid-structure interaction algorithm applied to the
  piston problem, Comp. Meth. Appl. Mech. Engrg. 167 (1998) 369--391.

\bibitem{hubner}
B.~H{\"u}bner, E.~Walhorn, D.~Dinkler, A monolithic approach to
  fluid--structure interaction using space--time finite elements, Comp. Meth.
  Appl. Mech. Engrg. 193~(23) (2004) 2087--2104.

\bibitem{turek2006}
J.~Hron, S.~Turek, A monolithic {FEM}/Multigrid solver for an {ALE} formulation
  of fluid-structure interaction with applications in Biomechanics, Springer,
  2006.

\bibitem{liu2014}
J.~Liu, R.~K. Jaiman, P.~S. Gurugubelli, A stable second-order scheme for
  fluid-structure interaction with strong added-mass effects, J. Comput. Phy.
  270 (2014) 687--710.

\bibitem{gurugubelli2015self}
P.~Gurugubelli, R.~Jaiman, Self-induced flapping dynamics of a flexible
  inverted foil in a uniform flow, Journal of Fluid Mechanics 781 (2015)
  657--694.

\bibitem{felippa01}
C.~A. Felippa, K.~C. Park, C.~Farhat, Partitioned analysis of coupled
  mechanical systems, Comp. Meth. Appl. Mech. Engrg. 190 (2001) 3247--3270.

\bibitem{cebral1997conservative}
J.~R. Cebral, R.~Lohner, Conservative load projection and tracking for
  fluid-structure problems, AIAA journal 35~(4) (1997) 687--692.

\bibitem{farhat05}
C.~Farhat, K.~G. van~der Zee, P.~Geuzaine, Provably second-order time-accurate
  loosely-coupled solution algorithms for transient nonlinear computational
  aeroelasticity, Comp. Meth. Appl. Mech. Engrg. 195 (2006) 1973--2001.

\bibitem{piperno_new}
S.~Piperno, C.~Farhat, Partitioned procedures for the transient solution of
  coupled aeroelastic problems part 1: Model problem, theory, and
  two-dimensional application, Comp. Meth. Appl. Mech. Engrg. 124 (1995)
  79--112.

\bibitem{yenduri2017new}
A.~Yenduri, R.~Ghoshal, R.~Jaiman, A new partitioned staggered scheme for
  flexible multibody interactions with strong inertial effects, Comp. Meth.
  Appl. Mech. Engrg. 315 (2017) 316--347.

\bibitem{jaiman_ficf2015}
R.~Jaiman, S.~Sen, P.~Gurugubelli, A fully implicit combined field scheme for
  freely vibrating square cylinders with sharp and rounded corners, Computers
  and Fluids 112 (2015) 1--18.

\bibitem{jaiman_cibc}
R.~K. Jaiman, P.~Geubelle, E.~Loth, X.~Jiao, Combined interface condition
  method for unsteady fluid-structure interaction, Comp. Meth. Appl. Mech.
  Engrg. 200 (2011) 27--39.

\bibitem{matthies}
H.~G. Matthies, R.~Niekamp, J.~Steindorf, Algorithms for strong coupling
  procedures, Comp. Meth. Appl. Mech. Engrg. 195 (2006) 2028--2049.

\bibitem{ahn}
H.~Ahn, Y.~Kallinderis, Strongly coupled flow/structure interactions with a
  geometrically conservative {ALE} scheme on general hybrid meshes, J. Comput.
  Phys. 219 (2006) 671--696.

\bibitem{jaiman2016stable}
R.~Jaiman, N.~Pillalamarri, M.~Guan, A stable second-order partitioned
  iterative scheme for freely vibrating low-mass bluff bodies in a uniform
  flow, Comp. Meth. Appl. Mech. Engrg. 301 (2016) 187--215.

\bibitem{farhat1998load}
C.~Farhat, M.~Lesoinne, P.~Le~Tallec, Load and motion transfer algorithms for
  fluid/structure interaction problems with non-matching discrete interfaces:
  Momentum and energy conservation, optimal discretization and application to
  aeroelasticity, Comp. Meth. Appl. Mech. Engrg. 157~(1-2) (1998) 95--114.

\bibitem{de2007review}
A.~De~Boer, A.~Van~Zuijlen, H.~Bijl, Review of coupling methods for
  non-matching meshes, Comp. Meth. Appl. Mech. Engrg. 196~(8) (2007)
  1515--1525.

\bibitem{jaiman2006conservative}
R.~K. Jaiman, X.~Jiao, P.~H. Geubelle, E.~Loth, Conservative load transfer
  along curved fluid--solid interface with non-matching meshes, Journal of
  Computational Physics 218~(1) (2006) 372--397.

\bibitem{jaiman_ijnme}
R.~Jaiman, X.~Jiao, X.~Geubelle, E.~Loth, Assessment of conservative load
  transfer on fluid-solid interface with nonmatching meshes, Int. J. Numer.
  Meth. Engng. 64 (2005) 2014--2038.

\bibitem{jiao2004common}
X.~Jiao, M.~T. Heath, Common-refinement-based data transfer between
  non-matching meshes in multiphysics simulations, International Journal for
  Numerical Methods in Engineering 61~(14) (2004) 2402--2427.

\bibitem{jiao2004overlaying}
X.~Jiao, M.~T. Heath, Overlaying surface meshes, part {I}: Algorithms,
  International Journal of Computational Geometry \& Applications 14~(06)
  (2004) 379--402.

\bibitem{slattery2016mesh}
S.~R. Slattery, Mesh-free data transfer algorithms for partitioned multiphysics
  problems: Conservation, accuracy, and parallelism, Journal of Computational
  Physics 307 (2016) 164--188.

\bibitem{jiao2004overlaying2}
X.~Jiao, M.~T. Heath, Overlaying surface meshes, part {II}: Topology
  preservation and feature matching, International Journal of Computational
  Geometry \& Applications 14~(06) (2004) 403--419.

\bibitem{brum1}
E.~van Brummelen, Added mass effects of compressible and incompressible flows
  in fluid-structure interaction, J. Appl. Mech. 76 (2009) 02106.

\bibitem{jaiman_jam}
R.~K. Jaiman., M.~K. Parmar, P.~S. Gurugubelli, Added mass and aeroelastic
  stability of a flexible plate interacting with mean flow in a confined
  channel, J. Appl. Mech. 81.

\bibitem{forster}
C.~Forster, W.~Wall, E.~Ramm, Artificial added mass instabilities in sequential
  staggered coupling of nonlinear structures and incompressible viscous flows,
  Comput. Methods Appl. Mech. Eng. 196 (2007) 1278--1293.

\bibitem{nobile}
P.~Causin, J.~F. Gerbeau, F.~Nobile, Added-mass effect in the design of
  partitioned algorithms for fluid-structure problems, Comput. Methods Appl.
  Mech. Eng. 194 (2005) 4506--4527.

\bibitem{jaiman2016partitioned}
R.~Jaiman, M.~Guan, T.~Miyanawala, Partitioned iterative and dynamic
  subgrid-scale methods for freely vibrating square-section structures at
  subcritical {R}eynolds number, Computers \& Fluids 133 (2016) 68--89.

\bibitem{hughes_ale}
T.~J.~R. Hughes, W.~Liu, T.~Zimmerman, {L}agrangian-{E}ulerian finite element
  formulation for incompressible visous flows, Comp. Meth. Appl. Mech. Engrg.
  29 (1981) 329--349.

\bibitem{donea}
J.~Donea, S.~Giuliani, J.~Halleux, Arbitrary {L}agrangian-{E}ulerian finite
  element method for transient dynamic fluid-structure interactions, Comp.
  Meth. Appl. Mech. Engrg. 33 (1982) 689--723.

\bibitem{bower2009applied}
A.~F. Bower, Applied mechanics of solids, CRC press, 2009.

\bibitem{jansen}
K.~Jansen, C.~Whitting, G.~Hulbert, A generalized-alpha method for integrating
  the filtered {N}avier-{S}tokes equations with a stabilized finite element
  method, Comp. Meth. Appl. Mech. Engrg. 190 (2000) 305--319.

\bibitem{yuri}
Y.~Bazilevs, K.~Takizawa, T.~Tezduar, Computational fluid-structure
  interaction: methods and aplications, Wiley, 2013.

\bibitem{antman_book}
S.~Antman, Nonlinear problems of elasticity, Springer-Verlag, New York, 2005.

\bibitem{breziniski2007}
C.~Breziniski, M.~Zaglia, Generalizations of {A}itken's process for
  accelerating the convergence of sequence, Journal of Computational and
  Applied Mathematics 26 (2007) 171--189.

\bibitem{buoso}
D.~Buoso, A.~Karapiperi, S.~Pozza, Generalizations of {A}itken's process for a
  certain class of sequences, Applied Numerical Mathematics 90 (2015) 38--54.

\bibitem{saad1986gmres}
Y.~Saad, M.~H. Schultz, Gmres: A generalized minimal residual algorithm for
  solving nonsymmetric linear systems, SIAM Journal on scientific and
  statistical computing 7~(3) (1986) 856--869.

\bibitem{woodsend2009hybrid}
K.~Woodsend, J.~Gondzio, Hybrid {MPI}/{O}pen{MP} parallel linear support vector
  machine training, Journal of Machine Learning Research 10~(Aug) (2009)
  1937--1953.

\bibitem{karypis1998software}
G.~Karypis, V.~Kumar, A software package for partitioning unstructured graphs,
  partitioning meshes, and computing fill-reducing orderings of sparse
  matrices, University of Minnesota, Department of Computer Science and
  Engineering, Army HPC Research Center, Minneapolis, MN.

\bibitem{smith2001development}
L.~Smith, M.~Bull, Development of mixed mode {MPI}/{O}pen{MP} applications,
  Scientific Programming 9~(2-3) (2001) 83--98.

\bibitem{turek2006proposal}
S.~Turek, J.~Hron, Proposal for numerical benchmarking of fluid-structure
  interaction between an elastic object and laminar incompressible flow, in:
  Fluid-structure interaction, Springer, 2006, pp. 371--385.

\bibitem{riserExperiment}
Vortex induced vibration data repository,
  \url{http://web.mit.edu.sg/towtank/www/vivdr/downloadpage.html}, datasets
  from ExxonMobil (Test case 1103).

\bibitem{vaibhav_CAF}
V.~Joshi, R.~K. Jaiman, A variationally bounded scheme for delayed detached
  eddy simulation: Application to vortex-induced vibration of offshore riser,
  Computers and Fluids (Under review),
  {\url{(https://www.dropbox.com/s/l7e77f0ikt8cqqr/CAF_Riser.pdf?dl=0)}}.

\end{thebibliography}
\end{document}